\documentclass[12pt]{article}
\usepackage{graphics}
\usepackage{a4wide}
\usepackage{amsfonts}
\usepackage{amstext}
\usepackage{amsmath}
\renewcommand{\theequation}{\arabic{section}.\arabic{equation}}
\newcommand{\Section}[1]{\setcounter{equation}{0}\section{#1}}

\def\d{\mbox{d}}

\def\ln{\mbox{ln}}

\def\2{I$\!$I}

\begin{document} 

\title{Density Profile of the
One-Dimensional Partially Asymmetric Simple Exclusion Process with
Open Boundaries}
\author{Tomohiro SASAMOTO \\
{\it Department of Physics, Graduate School of Science,}\\
{\it University of Tokyo,}\\
{\it Hongo 7-3-1, Bunkyo-ku, Tokyo 113-0033, Japan}}
\date{}

\maketitle

\begin{abstract}
The one-dimensional 
partially asymmetric simple 
exclusion process with open boundaries is considered.
The stationary state, which is known to be constructed 
in a matrix product form, is studied by applying the 
theory of $q$-orthogonal polynomials.
Using a formula of the $q$-Hermite polynomials,
the average density profile is computed in the 
thermodynamic limit.
The phase diagram for the correlation length, which was 
conjectured in \cite{me99}(J. Phys. A {\bf 32} (1999) 7109), 
is confirmed.

\vspace*{3mm}
\noindent
[Keywords: asymmetric simple exclusion, exact solution, 
density profile, $q$-orthogonal polynomials]
\end{abstract}

\Section{Introduction}
\label{intro}
The one-dimensional
asymmetric simple exclusion process (ASEP) \cite{L,Sp}
is a system of particles which hop preferentially in one direction
on a one-dimensional lattice with hard-core exclusion interaction.
The ASEP has been studied extensively since 
it is one of the few models which show rich non-equilibrium behaviors 
and is exactly solvable \cite{Derrida98}. 
Besides, the ASEP has applications to many interesting problems such as
the hopping conductivity, growth processes and the traffic flows \cite{SZ}.

In this article, we consider the stationary state of the ASEP 
with open boundary conditions.
That is, the system is connected to particle resevoirs at boundaries.
The case where particles can hop only in one direction, which 
we refer to as the ``totally asymmetric'' case in the sequel,
was solved in \cite{DEHP,SD}.
The current and the density profile were calculated 
exactly in the thermodynamic limit.
The phase diagram for the current and the correlation length
were identified.
The system exhibits phase transitions
depending on the parameters at the boundaries.
Recently the obtained phase diagram was discussed from the point of view of 
the domain wall dynamics \cite{KSKS}.

The partially asymmetric case with 
the open boundary conditions was partially solved in \cite{me99}. 
The current was evaluated in the thermodynamic limit.
The phase diagram for the current was identified.
It turned out to be the same as the one 
obtained by mean-field approximation \cite{ER} or
by employing a plausible assumption \cite{Sandow}.
The phase diagram for the correlaiton length was 
also obtained by assuming that the correlation length
is given by the logarithm of the ratio of the largest and
the second largest eigenvalue of a certain matrix
which plays a similar role as a transfer matrix does
in equilibrium statistical mechanical models.
It was shown that the phase diagram has a richer structure than 
that for the totally asymmetric case.
The average density profile was, however, not calculated in \cite{me99}.
In this sence the obtained phase diagram for the correlation length
has remained a conjecture.
The purpose of this paper is the confirmation of this phase diagram.
By using the explicit formula for the Poisson kernel of the $q$-Hermite 
polynomials, the average density profile in the thermodynamic limit
is calculated for 
the partially asymmetric case.
It turns out that the phase diagram was correctly predicted 
in \cite{me99}.

In this article, 
we only consider the case where hoppings of particles 
at the boundaries and those at the bulk part of the system are compatible. 
In other words, when we allow the particle input at the left 
boundary and the particle output at the right boundary,
the hopping rate to the right is assumed to be larger than 
that to the left. When hoppings at the boundaries and 
those at bulk is imcompatible, the current becomes zero 
in the thermodynamic limit.
The situation seems to be similar to the closed boundary
condition where particles can not enter or go out of the system
\cite{SS}.
Off course, when we consider the finite chain, the current remains to 
be positive.
We remark that the asymptotic current for this case was
evaluated in \cite{BECE}.

The paper is organized as follows.
In the next section, the definition of the model 
is given in terms of the master equation.
The so-called matrix product ansatz, which gives the 
stationary state in the form of matrix product, is also explained.
Some properties of the $q$-Hermite polynomials and the 
relationship to the matrix product ansatz 
are explained in section \ref{q-H}.
The section \ref{density} is the main section of this article.
First, the one-point funciton is
represented in the form of double integrals.
Second, the average density profile in the thermodynamic limit
are summerlized whereas the evaluation of the
integrals are relegated to Appendices. 
The phase diagram for the correlation length is 
identified.
The concluding remarks are given in the last section.

\Section{Definition of Model and Matrix Product Ansatz}
The one-dimensional asymmetric simple exclusion process (ASEP)
is defined as follows.
During the infinitesimal time interval $\d t$, 
each particle jumps to the 
right nearest neighboring site with probability 
$p_R \d t$ and to the left 
nearest neighboring site with probability $p_L \d t$.
If the chosen site is already occupied, the particle 
does not move due to the exclusion rule.
More than one particle can 
not be on the same site.
Each site can be either empty or
occupied.
The case where particles can hop only in one direction, 
i.e., the case where either $p_L = 0$ or $p_R=0$ is called the 
``totally asymmetric'' case. 
The $p_R=p_L$ case is called the ``symmetric'' case whereas 
the case where particles hop in both
directions with different rates will be referred to 
as the ``partially asymmetric'' case.
In addition, we allow the particle input at the left end of the chain
with rate $\alpha$ and allow the particle output 
at the right end of the chain with rate $\beta$
(Fig. 1).
Here the length of the chain is denoted by $L$. 
In this article, we restrict our attention to 
the partially asymmetric case since the totally asymmetric case 
and the symmetric case was already solved in \cite{DEHP,SD} and in 
\cite{me96} respectively. 
The restrictions on the parameters are
$0 < p_L < p_R$ and $\alpha,\beta>0$.

More formally, the process is defined in terms of the master equation.
Each configuration of the system is indicated by 
$\{\tau_1,\tau_2,\ldots,\tau_L\}$ where 
$\tau_j$ $(j=1,2,\ldots,L)$ denotes the particle number
at site $j$. Namely $\tau_j=0$ if the site $j$ is empty
whereas $\tau_j=1$ if the site $j$ is occupied.  
Let $P(\tau_1,\tau_2,\ldots,\tau_L;t)$ denote
the probability that the system has the configulation 
$\{ \tau_1,\tau_2,\ldots,\tau_L\}$ at time $t$.
Then the  time evolution of the ASEP is described by the 
following master equation,
\begin{align}
  &\quad
  \frac{\d}{\d t} P(\tau_1,\tau_2,\ldots,\tau_L;t)
  \notag
  \\
  &=
  \alpha (2\tau_1-1) P(0,\tau_2,\ldots,\tau_L;t)
  \notag
  \\
  &\quad 
  +
  \sum_{j=1}^{L-1}
  (\tau_j-\tau_{j+1})
  \left[ 
   p_L    P(\tau_1,\tau_2,\ldots,0,1,\ldots,\tau_L;t)
  \right. 
  \notag
  \\
  &\quad
  \left.
   -
    p_R    P(\tau_1,\tau_2,\ldots,1,0,\ldots,\tau_L;t) 
  \right]
  \notag
  \\
  &\quad
  +
  \beta (1-2\tau_L) P(\tau_1,\tau_2,\ldots,\tau_{L-1},1;t). 
  \label{mas-eq}
\end{align}
For instance, the master equation for $L=2$ case reads
\begin{equation}
\label{mas-eq-L2}
\frac{\d}{\d t} 
\begin{bmatrix}
P(00;t)\\
P(01;t)\\
P(10;t)\\
P(11;t)
\end{bmatrix}
=
-
\begin{bmatrix}
\alpha  & -\beta              &  0   & 0\\
0       & \alpha + p_L +\beta & -p_R & 0\\ 
-\alpha & -p_L                &  p_R & -\beta\\
0       & -\alpha             &  0   & \beta
\end{bmatrix}
\begin{bmatrix}
P(00;t)\\
P(01;t)\\
P(10;t)\\
P(11;t)
\end{bmatrix}.
\end{equation}
One can confirm himself that the dynamics 
of the ASEP is correctly encoded in the master equation
(\ref{mas-eq}). 

When time $t$ goes to infinity, the system is expected to 
reach the stationary state.
The probability distribution in the stationary state will be 
denoted as $P(\tau_1,\tau_2,\ldots,\tau_L)$. 
For instance, except for the normalization, 
the stationary state for $L=2$ case is 
the eigenvector of the $4\times 4$ matrix in the right 
hand side of (\ref{mas-eq-L2}) with the eigenvalue zero.
Explicitly, it reads
\begin{equation}
\label{stationary-2}
\begin{bmatrix}
P(00)\\
P(01)\\
P(10)\\
P(11)
\end{bmatrix}
=
\text{Const.}
\begin{bmatrix}
\frac{1}{\alpha^2} \\
\frac{1}{\alpha\beta}\\
\frac{1}{p_R}
\left( 
\frac{p_L}{\alpha\beta} + \frac{1}{\alpha} +\frac{1}{\beta} 
\right)
\\
\frac{1}{\beta^2}  
\end{bmatrix}.
\end{equation}

In \cite{DEHP}, it was shown that 
the probability distribution of the ASEP in the stationary state 
for general $L$ can be written in the form of the matrix product as 
\begin{equation}
  \label{originalMPA}
  P(\tau_1,\tau_2,\ldots,\tau_L) 
  =
  \frac{1}{Z_L}
  \langle W| \prod_{j=1}^{L}(\tau_j D + (1-\tau_j) E)| V\rangle,   
\end{equation}
where $D$ and $E$ are square matrices and 
$\langle W|$ and $|V\rangle$ are vectors satisfying
following relations, 
\begin{subequations}
\begin{gather}
  \label{mat-cond1}
  p_R DE-p_L ED 
  = 
  \zeta (D+E),
  \\
  \label{mat-cond2}
  \alpha \langle W| E 
  = 
  \zeta\langle W|,
  \quad
  \beta D |V\rangle 
  = 
  \zeta |V\rangle.
\end{gather}
\end{subequations}
Here $\zeta$ is an arbitrary number.
If one defines the 
matrix $C$ by
\begin{equation}
  \label{eq:def-c}
C=D+E,
\end{equation}
the normalization $Z_L$ is given by
\begin{equation}
Z_L
=
\langle W| C^L |V\rangle. 
\end{equation}

Here, for the case of $L=2$, we check that the 
state (\ref{originalMPA}) indeed gives the stationary state 
of the process by using the algebraic relations 
(\ref{mat-cond1}) and (\ref{mat-cond2}).
For $L=2$, (\ref{originalMPA}) reads
\begin{equation}
\begin{bmatrix}
P(00) \\
P(01) \\
P(10) \\
P(11) 
\end{bmatrix}
=
\frac{1}{Z_2}
\begin{bmatrix}
\langle W| E^2 |V\rangle \\
\langle W| ED  |V\rangle \\
\langle W| DE  |V\rangle \\
\langle W| D^2 |V\rangle 
\end{bmatrix}.
\end{equation}
Three components, $P(00),P(01),P(11)$ can be calculated by 
simply using (\ref{mat-cond2}). 
On the other hand, one computes $P(10)$  
first changing the order of matrices $D,E$ by  
(\ref{mat-cond1}) and then using (\ref{mat-cond2}).
Hence we get
\begin{equation}
\begin{bmatrix}
P(00) \\
P(01) \\
P(10) \\
P(11) 
\end{bmatrix}
=
\frac{1}{Z_2}
\begin{bmatrix}
\frac{\zeta^2}{\alpha^2}\\
\frac{\zeta^2}{\alpha\beta}\\
\frac{\zeta^2}{p_R}
\left( \frac{p_L}{\alpha\beta} + \frac{1}{\alpha} +\frac{1}{\beta} 
\right)\\
\frac{\zeta^2}{\beta^2}
\end{bmatrix}.
\end{equation}
One can compare this expression with (\ref{stationary-2})
to see that this expression indeed gives the stationary 
state for $L=2$ case.
One also sees that the arbitrary parameter $\zeta$  
appears in the same way for all components,
$P(00),P(01),P(10),P(11)$. Changing the parameter $\zeta$ 
only changes the normalization $Z_2$.
This is true for general $L$ as well. 
In this article, the proof that the state (\ref{originalMPA})
gives the stationary state for general $L$ 
is not given. See \cite{DEHP}.  

We express several physical quantities 
in the form of matrix products.
The one-point function $\langle n_j \rangle_L$ is
defined as the probability that the site $j$ is
occupied.
In other words, $\langle n_j \rangle_L$ is 
the average density at site $j$. 
The two-point function $\langle n_j n_k\rangle_L$ 
is defined as the probability that the sites $j$ 
and the site $k$ are both occupied. 
Higher correlation functions are defined similarly.
In the matrix language, they are computed by
\begin{align}
\label{def-1pt}
\langle n_j \rangle_L
&=
\langle W| C^{j-1} D C^{L-j} |V\rangle /Z_L ,
\\
\label{def-2pt}
\langle n_j n_k\rangle_L
&=
\langle W| C^{j-1} D C^{k-j-1} D C^{L-k}|V\rangle /Z_L ,
\end{align}
and so on.
The current through the bond between 
site $j$ and site $j+1$ is defined by 
$
J_L^{(j)} 
= 
p_R \langle n_j (1-n_{j+1})\rangle
-
p_L \langle n_j (1-n_{j-1})\rangle.
$
In the steady state, the current is independent of $j$ 
and hence is denoted by $J_L$.
It is given by
\begin{equation}
  \label{def-current}
J_L
=
\zeta
\frac{\langle W|C^{L-1}|V\rangle}
     {\langle W|C^{L}|V\rangle}
=
\zeta 
\frac{Z_{L-1}}
     {Z_L}.
\end{equation}
Once one finds a representation of these algebraic relations, 
by using the above formula,
one can in principle calculate the physical 
quantities such as the particle current $J_L$,
the one-point function $\langle n_j \rangle_L$,
the two-point function $\langle n_j n_k\rangle_L$
and the higher correlation functions.

We note that the process has 
an obvious particle-hole symmetry.
When we look at holes instead of particles, they
tend to hop to the left with rate $p_R$ and to the 
right with rate $p_L$ with hard-core exclusion. 
In addition, they are
injected at right end with rate $\beta$ and they are
removed at the left end with rate $\alpha$.
In other words, the process is invariant under the
changes,
\begin{align}
\notag
\text{particle}
&\leftrightarrow
\text{hole}
\\
\label{symmetry}
\alpha
&\leftrightarrow 
\beta
\\
\notag
\text{site number} \,\, j
&\leftrightarrow 
\text{site number} \,\, L-j+1.
\end{align}
Due to this symmetry, it is sufficient to obtain
the density for the right half of the system.
The density for the left half of the system is obtained by
using the above symmetry as
\begin{equation}
  \label{ri-le}
\langle n_j \rangle_L (\alpha,\beta)
=
1-\langle n_{L-j+1} \rangle_L (\beta,\alpha),  
\end{equation}
where the dependence of $\langle n_j \rangle_L$ 
on the parameters $\alpha$ and $\beta$ are 
eplicitly indicated.

Before closing the section, we present some simulation
results (Fig. 2 and Fig. 3). 
Figure 2 shows the space-time 
diagrams for several choices of parameters.
It is clear that the properties of the system
crutially depend on the values of the boundary 
parameters. 
The differeces become more transparent
when we consider the particle current or the 
the average density profile of the stationary state.
In principle, the stationary state is achieved only in the
infinite time limit. However,
we see from Fig. 2 that the system practically
goes into a stationary state after some transient time.
Hence if we average the density over a long time 
after the transient time, it would be regarded as
the average density profile of the stationary state practically.
The results are shown in Fig. 3.
When $\alpha$ is small and $\beta$  is large,
the bulk density is low. It decays sharply near
the right boundary. This is called the low density phase.
Conversely, when $\alpha$ is large and $\beta$  is small,
the bulk density is high. It decays sharply near
the left boundary. This is called the high density phase.
When $\alpha=\beta$ is small, the low density region 
and the high density region coexist (coexistence line).
Finally, when both $\alpha$ and $\beta$ are 
large enough, the density takes the value $1/2$ 
at bulk and decays slowly near both the boundaries.
This is called the maximal current phase.
Our main tasks in the following are to obtain the average density profiles
in Fig. 4 exactly.

\Section{Representation of Algebra and $q$-Hermite Polynomials}
\label{q-H}
First we introduce some notations 
for later convenience.
We introduce the $q$-number, 
\begin{equation}
\{n\}
=
1-q^n,
\end{equation}
and the 
$q$-shifted factorial,
\begin{subequations}
\begin{align}
  (a;q)_n
  &=
  \label{q-shi-fac-1}
  (1-a)(1-aq)(1-aq^2)\cdots(1-aq^{n-1}),
  \\
  (a;q)_0
  &=
  1.
  \label{q-shi-fac-2}
\end{align}
\end{subequations}
We also define 
\begin{equation}
  \label{q-prod-inf}
  (a;q)_{\infty}
  =
  \prod_{j=0}^{\infty}(1-aq^j),
\end{equation}
for $|q|<1$.
Since products of $q$-shifted factorials appear so often,
we use the notations,
\begin{align}
(a_1,a_2,\cdots,a_k;q)_{\infty}
&=
(a_1;q)_{\infty}(a_2;q)_{\infty}\cdots (a_k;q)_{\infty},
\\
(a_1,a_2,\cdots,a_k;q)_{n}
&=
(a_1;q)_{n}(a_2;q)_{n}\cdots (a_k;q)_{n} .
\end{align}

Next a representation of the algebraic relaiton
(\ref{mat-cond1}) and (\ref{mat-cond2}) is given.
If we define
\begin{equation}
D
=
1+d,
\hspace{10mm}
E
=
1+e,
\end{equation}
we see that these relations become 
\begin{subequations}
\begin{gather}
\label{q-b}
d e - q e d 
=
1-q,
\\
\label{w-coh}
\langle W| e 
=
a \langle W|,
\hspace{10mm}
d |V\rangle
=
b |V\rangle,
\end{gather}
\end{subequations}
where we put
\begin{gather}
q\
=
p_L/p_R,
\\
a
=
\frac{1-\tilde{\alpha}}
     {\tilde{\alpha}},
\quad
b
=
\frac{1-\tilde{\beta}}
     {\tilde{\beta}},
\end{gather}
with
$\tilde{\alpha}=\alpha/(p_R-p_L),
\tilde{\beta}=\beta/(p_R-p_L)$.
Since $0< p_R<p_L$, 
we have $0<q<1$.

In this article, we take the 
following representation for the matrices $d, e$ 
and the vectors $\langle W|, |V\rangle$,
\begin{subequations}
\begin{gather}
\label{de-qb}
d
=
\begin{bmatrix}
0 & \{1\}^{\frac{1}{2}} & 0                   & 0                & \cdots  
 \\
0       & 0             & \{2\}^{\frac{1}{2}} & 0                &     \\
0       & 0             & 0                   & \{3\}^{\frac{1}{2}} &  \\
\vdots  &               &                     & \ddots           &\ddots \\
\end{bmatrix},
\hspace{10mm}
e
=
\begin{bmatrix}
0      & 0 & 0 & 0 & \cdots \\
\{1\}^{\frac{1}{2}} & 0 & 0 & 0 &   \\
0      & \{2\}^{\frac{1}{2}} & 0 & 0 &   \\
0      & 0 & \{3\}^{\frac{1}{2}} & 0 &   \\
\vdots &   & & \ddots & \ddots
\end{bmatrix},
\\
\label{WV-qb}
\langle W| 
= 
\kappa\,_c\langle a|
=
\kappa \left( 1, \frac{a}{\sqrt{(q;q)_1} },
             \frac{a^2}{ \sqrt{(q;q)_2} },\ldots \right),
\hspace{10mm}
|V \rangle 
= 
\kappa \, |b\rangle_c
=
\kappa 
\left(
\begin{matrix}
1\\
\displaystyle\frac{b}{ \sqrt{(q;q)_1}}\\
\displaystyle\frac{b^2}{ \sqrt{(q;q)_2} }\\
\vdots
\end{matrix}
\right).
\end{gather}
\end{subequations}
The constant $\kappa$ is takes as $\kappa^2=(ab;q)_{\infty}$
so that $\langle W|V\rangle=1$.
It should be noticed that there exists another 
useful representaion of the algebraic 
relations (\ref{mat-cond1}) and (\ref{mat-cond2}).
It was first given in \cite{DEHP} and was used to 
obtain the phase diagram of the correlation length 
in \cite{me99}.
The advantage of the represetation (\ref{de-qb}) and 
(\ref{WV-qb}) is that the commutation relation 
of the matrices $d$ and $e$ turns out to be a simple
diagonal matrix. We have
\begin{equation}
  \label{de-com}
  de-ed
  =
(1-q)
\begin{bmatrix}
1      & 0 & 0   & 0      & \cdots \\
0      & q & 0   & 0      &        \\
0      & 0 & q^2 & 0      &        \\
0      & 0 & 0   & q^3    &        \\
\vdots &   &     &        & \ddots
\end{bmatrix}.
\end{equation}
This fact will play an important role for the calculation of the 
average density profile in the next section.

Next we list some properties of the 
continuous  $q$-Hermite polynomials .
The proofs can be found for instance in \cite{AAR,GR}.
The continuous  $q$-Hermite polynomials 
$\{ H_n(x|q) |\, n=0,1,2,\ldots\}$ are defined by 
the three term recurrence relation,
\begin{equation}
\label{rec-qh}
 H_{n+1}(x;q)
+
(1-q^n)H_{n-1}(x;q)
=
2 x H_n(x;q),
\end{equation}
with the initial condition,
\begin{equation}
  \label{init-cond}
H_{-1}(x;q)
=
0,
\hspace{10mm}
H_0(x;q)
=
1.
\end{equation}
They are explicitly given by
the formula,
\begin{equation}
H_n(\cos\theta|q)
=
\sum_{k=0}^{n}
\frac{(q;q)_n }{ (q;q)_k (q;q)_{n-k} }
e^{i(n-2k)\theta}.
\end{equation} 
The orthogonality relation reads
\begin{equation}
\label{ortho}
\int_{0}^{\pi}
H_n(\cos\theta|q) H_m(\cos\theta|q) 
(e^{2i\theta},e^{-2i\theta};q)_{\infty}\d \theta
=
2\pi \frac{ (q;q)_n }{ (q;q)_{\infty} } \delta_{mn}. 
\end{equation}
The generating function is also known and is given by 
\begin{equation}
\label{gen}
\sum_{n=0}^{\infty}
\frac{ H_n(\cos\theta|q) }{ (q;q)_n }
\lambda^n
=
\frac{1}{(\lambda e^{i\theta},\lambda e^{-i\theta};q)_{\infty} }
\end{equation}
for $|\lambda| < 1$.
To calculate the average density profile, 
we also need the so-called Poisson kernel,
\begin{equation}
  \label{Mehler}
  \sum_{n=0}^{\infty}
  \frac{ H_n(\cos\theta|q) H_n(\cos\varphi|q) r^n }
       {(q;q)_n}
  =
  \frac{ (r^2;q)_{\infty} }
       { (r e^{i(\theta+\varphi)}, r e^{-i(\theta+\varphi)}, 
          r e^{i(\theta-\varphi)}, r e^{-i(\theta-\varphi)} ;q)_{\infty} }.
\end{equation}
This formula is called $q$-Mehler formula in the mathematics literature.

Here we notince that,
if we introduce 
\begin{equation}
  \label{p-Hermite}
  p_n(x)
  =
  H_n(x|q)/\sqrt{(q;q)_n},
\end{equation}
the three-term recurrence relation is rewritten into the form,
\begin{equation}
\label{p-eigen}
\begin{bmatrix}
0 & \{1\}^{\frac{1}{2}} & 0  & 0  & \cdots    \\
\{1\}^{\frac{1}{2}} & 0  & \{2\}^{\frac{1}{2}} & 0 &     \\
0 & \{2\}^{\frac{1}{2}} & 0  & \{3\}^{\frac{1}{2}} &  \\
\vdots  & & \ddots & \ddots           &\ddots \\
\end{bmatrix}
\begin{bmatrix}
p_ 0(x)\\
p_ 1(x)\\
p_ 2(x)\\
\vdots 
\end{bmatrix}
=
2x 
\begin{bmatrix}
p_ 0(x)\\
p_ 1(x)\\
p_ 2(x)\\
\vdots 
\end{bmatrix}.
\end{equation}
In other words, $|p(x)\rangle =\,^t(p_0(x), p_1(x), \ldots)$
is formally an eigenvector of the matrix $d+e$ with 
eigenvalue $2x$.
This is the basic relationship between the representaion 
of the algebra (\ref{mat-cond1}),(\ref{mat-cond2}) and 
the theory of $q$-orthogonal polynomials.
Finally, the completeness of the continuous $q$-Hermite 
polynomials reads 
\begin{equation}
  \label{complete}
1
=
\frac{(q;q)_{\infty}}
     {2\pi}
\int_0^{\pi}
\d \theta (e^{i\theta},e^{-i\theta};q)_{\infty}
|p(\cos\theta)\rangle 
\langle p(\cos\theta) |.
\end{equation}

\Section{Calculation of Density Profile}
\label{density}
In this section, the average density profile is calculated by 
using the formula (\ref{Mehler}) of the continuous
$q$-Hermite polynomials.
We first recall the asymptotic behaviors of 
the normalization $Z_L$ and the current 
in the thermodynamic limit $J=\lim_{L\rightarrow\infty} J_L$.
The asymptotic expressions for $Z_L$ were given in \cite{me99}
and are summarized as follows:
\begin{itemize}
  \item For phase $A$
        (low-density phase; $a>1$ and $a>b$ ;
         $\tilde{\alpha} < \frac12$ 
         and $\tilde{\alpha} <  \tilde{\beta}$)
  \begin{equation}
  Z_L 
  \simeq
  \frac{(a^{-2};q)_{\infty}}
       {(b/a;q)_{\infty}}
  [(1+a)(1+a^{-1})]^L ,
  \end{equation}

  \item For phase $B$
        (high-density phase; $b>1$ and $a<b$ ;
         $\tilde{\beta} < \frac12$ 
         and $\tilde{\alpha} >  \tilde{\beta}$)
  \begin{equation}
  Z_L 
  \simeq
  \frac{(b^{-2};q)_{\infty}}
       {(a/b;q)_{\infty}}
  [(1+b)(1+b^{-1})]^L ,
  \end{equation}

  \item For phase $C$ 
        (maximal current phase; $0<a,b<1$ ;
         $\tilde{\beta} > \frac12$ 
         and $\tilde{\alpha} > \frac12$)
  \begin{equation}
  Z_L
  \simeq
  \frac{(ab;q)_{\infty}  (q;q)_{\infty}^3 4^{L+1}}
       {\sqrt{\pi}(a,b;q)_{\infty}^2 
        L^{ \frac{3}{2} }            } ,
  \end{equation}

  \item On the coexistense line
        ($a=b>1$ ; $\tilde{\alpha} = \tilde{\beta} < \frac12$)
  \begin{equation}
  Z_L 
  \simeq
  \frac{(a-a^{-1}) (a^{-2};q)_{\infty} L}
       {(q;q)_{\infty}}
  [(1+a)(1+a^{-1})]^{L-1}.
  \end{equation}
\end{itemize}

Using (\ref{def-current}), 
the current in the thermodynamic limit is readily computed as 
\begin{itemize}
\item For phase A
      ($\tilde{\alpha} < \frac12$ 
       and $\tilde{\beta} >\tilde{\alpha}$)
  \begin{equation}
    \label{current-A}
    J=(p_R-p_L)\tilde{\alpha} (1-\tilde{\alpha}),
  \end{equation}
\item For phase B
      ($\tilde{\beta} < \frac12$ 
       and $\tilde{\alpha} >\tilde{\beta}$)
  \begin{equation}
    \label{current-B}
    J=(p_R-p_L)\tilde{\beta} (1-\tilde{\beta}),
  \end{equation}
\item For phase C  
      ($\tilde{\alpha} > \frac12$ 
       and $\tilde{\beta} >\frac12$)
  \begin{equation}
    \label{current-C}
    J=\frac{p_R-p_L}{4}.
  \end{equation}
\end{itemize}
The phase diagram for the current is depicted in Fig. 4.

Now we turn to consider the average density profile.
As you can see from the figures in Fig. 4, 
the average density is almost 
constant at bulk part except on the coexistence line. 
Hence we are interested in the average bulk density and 
how the density decays near the boundaries.
When the average density decays like $e^{-r/\xi}$ with 
$r$ distance from the boundary, we refer to $\xi$ as 
the correlation length in this paper.
As for the average density near the boundares,
it is sufficient to compute the density 
near the right boundary
due to the symmetry (\ref{symmetry}).
The relation (\ref{ri-le}) enables us to know the
average density profile near the left boundary.
On the other hand,
the coexistence line should be treated separately.
Since it is easier 
to calculate the density difference than the density itself,
we rewrite the density at site $j$ as
\begin{equation}
  \label{density-decompose}
  \langle n_j \rangle_L
  =
  \sum_{k=j}^{L-1}
  (\langle n_k \rangle_L - \langle n_{k+1} \rangle_L)
  +
  \langle n_L \rangle_L .
\end{equation}
At the right boundary, we have
\begin{align}
  \label{d-right}
  \langle n_L \rangle_L
  &=
  \frac{1}{Z_L}
  \langle W| C^{L-1} D |V\rangle 
  \notag\\
  &=
  \frac{1}{\tilde{\beta}} \frac{Z_{L-1}}{Z_L}  
  \notag\\
  &\rightarrow
  \frac{J}{\beta}.
  \hspace{10mm}
  (L\rightarrow\infty)
\end{align}
Using (\ref{current-A})-(\ref{current-C}), 
the density at the right boundary is easily calculated.

Next we notice that
\begin{align}
  \langle n_k \rangle_L
  -
  \langle n_{k+1} \rangle_L
  &=
  \frac{1}{Z_L}
  (\langle W| C^{k-1}DC^{L-k}|V\rangle
   -
   \langle W| C^{k}DC^{L-k-1}|V\rangle )
  \notag\\
  &=
  \frac{1}{Z_L}
  \langle W|C^{k-1}(DC-CD)C^{L-k-1}|V\rangle
  \notag\\
  &=
  \frac{1}{Z_L}
  \langle W|C^{k-1}(DE-ED)C^{L-k-1}|V\rangle
  \notag\\
  &=
  \frac{1}{Z_L}
  \langle W|C^{k-1}(de-ed)C^{L-k-1}|V\rangle.
\end{align}
Now one can represent $\langle W|C^{k-1}(de-ed)C^{L-k-1}|V\rangle$ 
in the form of double integrals.
The calculation proceeds as follows.
First we notice that 
\begin{align}
  &\quad
  \langle W|C^{k-1}(de-ed)C^{L-k-1}|V\rangle
  \notag\\
  &=
  \kappa^2 
  (q;q)_{\infty}^2
  \int_0^{\pi}
  \frac{\d \theta}{2\pi}
  (e^{2i\theta},e^{-2i\theta};q)_{\infty}
  \int_0^{\pi}
  \frac{\d \varphi}{2\pi}
  (e^{2i\varphi},e^{-2i\varphi};q)_{\infty}
  \notag\\
  &\quad \times \,
  _c \langle a|
  C^{k-1}
  |p(\cos\theta)\rangle 
  \langle p(\cos\theta)|
  (de-ed)
  C^{L-k-1}
  |p(\cos\varphi)\rangle
  \langle p(\cos\varphi)| b\rangle_c
  \notag\\
  &=
  (ab;q)_{\infty}
  (q;q)_{\infty}^2
  \int_0^{\pi}
  \frac{\d \theta}{2\pi}
  \int_0^{\pi}
  \frac{\d \varphi}{2\pi}
  (e^{2i\theta},e^{-2i\theta},e^{2i\varphi},e^{-2i\varphi};q)_{\infty}
  [2(1+\cos\theta)]^{k-1}[2(1+\cos\varphi)]^{L-k-1}  
  \notag\\
  &\quad\times \,
  _c \langle a| p(\cos\theta)\rangle 
  \langle p(\cos\theta)|
  (de-ed)
  |p(\cos\varphi)\rangle
  \langle p(\cos\varphi)| b\rangle_c .    
\end{align}
Here the formula (\ref{gen}) gives
\begin{equation}
  \label{ap-pb}
_c \langle a| p(\cos\theta)\rangle   
=
\frac{1}{(ae^{i\theta},a^{-i\theta};q)_{\infty}},
\quad
\langle p(\cos\varphi)| b\rangle_c 
=
\frac{1}{(be^{i\theta},b^{-i\theta};q)_{\infty}}.    
\end{equation}
The remaining term 
$ \langle p(\cos\theta)|
  (de-ed)
  |p(\cos\varphi)\rangle
$ can also be computed by using 
the fact (\ref{de-com}) and the formula (\ref{Mehler}).
We see that
\begin{align}
\langle p(\cos\theta)|
(de-ed)
|p(\cos\varphi)\rangle  
&=
(1-q)
\sum_{n=0}^{\infty}
p_n(\cos\theta) p_n(\cos\varphi) q^n
\notag\\
&=
(1-q)
\sum_{n=0}^{\infty}
  \frac{ H_n(\cos\theta|q) H_n(\cos\varphi|q) q^n }
       {(q;q)_n}
\notag\\
&=
  \frac{ (q;q)_{\infty} }
       { (q e^{i(\theta+\varphi)}, q e^{-i(\theta+\varphi)}, 
          q e^{i(\theta-\varphi)}, q e^{-i(\theta-\varphi)} ;q)_{\infty} }
\end{align}
Hence we have
\begin{gather}
\langle W|C^{k-1}(de-ed)C^{L-k-1}|V\rangle
\notag\\
=   
(ab;q)_{\infty}(q;q)_{\infty}^3
\int_0^{\pi}
\frac{\d \theta}{2\pi}
\int_0^{\pi}
\frac{\d \varphi}{2\pi}
\frac{
      (e^{2i\theta},e^{-2i\theta},e^{2i\varphi},e^{-2i\varphi};q)_{\infty}
      [2(1+\cos\theta)]^{k-1}[2(1+\cos\varphi)]^{L-k-1}  
     }
     {(ae^{i\theta},        ae^{-i\theta},
       qe^{i(\theta+\varphi)}, qe^{-i(\theta+\varphi)}, 
       qe^{i(\theta-\varphi)}, qe^{-i(\theta-\varphi)}, 
       be^{i\varphi},        be^{-i\varphi};q)_{\infty} 
     }
\end{gather}
for $0<a,b<1$.
Summing with respect to $k$ from $j$ to $L-1$, we get
\begin{gather}
  \sum_{k=j}^{L-1}
  (\langle n_k \rangle_L - \langle n_{k+1} \rangle_L)  
  \notag\\
  =
  \frac{(ab;q)_{\infty}(q;q)_{\infty}^3}{Z_L}
  \int_0^{\pi}
  \frac{\d \theta}{2\pi}
  \int_0^{\pi}
  \frac{\d \varphi}{2\pi}
  \frac{
        (e^{2i\theta},e^{-2i\theta},e^{2i\varphi},e^{-2i\varphi};q)_{\infty}
       }
       {(ae^{i\theta},        ae^{-i\theta},
         qe^{i(\theta+\varphi)}, qe^{-i(\theta+\varphi)}, 
         qe^{i(\theta-\varphi)}, qe^{-i(\theta-\varphi)}, 
         be^{i\varphi},        be^{-i\varphi};q)_{\infty} 
       }
  \notag\\
  \times
  \frac{
           [2(1+\cos\theta)]^{j-1}[2(1+\cos\varphi)]^{L-j} 
          -[2(1+\cos\theta)]^{L-1} 
       }
       { 2\cos\varphi - 2 \cos\theta },
  \label{diff}
\end{gather}
 when $0<a,b<1$.
Hence, substitution of this expression into (\ref{density-decompose})
leads to the expression of the average density profile for $0<a,b<1$.
Finally, the analytic continuation of (\ref{diff}) 
gives the expression of the density porfile 
for other values of the parameters as well.
This is conveniently done by writing (\ref{diff}) as a contour 
integral of the two complex variables $z_1=e^{i\theta}$ and 
$z_2=e^{i\varphi}$.
That is, we rewrite (\ref{diff}) as
\begin{equation}
  \label{I1-I2}
  \sum_{k=j}^{L-1}
  (\langle n_k \rangle_L - \langle n_{k+1} \rangle_L)  
  =
  \frac{1}{Z_L}(I_1+I_2),  
\end{equation}
where
\begin{align}
I_1
&=
\frac14 
(ab;q)_{\infty}(q;q)_{\infty}^3
\int\frac{\d z_1}{2\pi i z_1}
\int\frac{\d z_2}{2\pi i z_2}
\notag\\
&\quad
\frac{(z_1^2,z_1^{-2},z_2^2,z_2^{-2};q)_{\infty}
      [(1+z_1)(1+z_1^{-1})]^{L-1}}
     {(az_1,a z_1^{-1},qz_1 z_2, q z_1^{-1} z_2^{-1},
       q z_1 z_2^{-1},q z_1^{-1} z_2^{-1}, bz_2,b z_2^{-1};q)_{\infty}
      (z_2 + z_2^{-1} -z_1-z_1^{-1})},
\notag\\
I_2
&=
\frac14 
(ab;q)_{\infty}(q;q)_{\infty}^3
\int\frac{\d z_1}{2\pi i z_1}
\int\frac{\d z_2}{2\pi i z_2}
\notag\\
&\quad
\frac{(z_1^2,z_1^{-2},z_2^2,z_2^{-2};q)_{\infty}
      [(1+z_1)(1+z_1^{-1})]^{j-1}
      [(1+z_2)(1+z_2^{-1})]^{L-j} }
     {(az_1,a z_1^{-1},q z_1 z_2, q z_1^{-1} z_2^{-1},
       q z_1 z_2^{-1},q z_1^{-1} z_2^{-1}, bz_2,b z_2^{-1};q)_{\infty}
      (z_2 + z_2^{-1} -z_1-z_1^{-1})}.
\end{align}
When $0<a,b<1$, the contours of  $z_1$ and $z_2$ are both unit circles.
For other valures of the parameters, the contours are deformed  
as poles in the integrands move in and out of the unit circles.

The evaluation of the integrals $I_1$ and $I_2$ in the 
thermodynamic limit for each phase is relegated to  
Appendices.
Basically, the integral $I_1$ gives the density at bulk region 
whereas the integral $I_2$ gives the density near the right boundary.
The results are summerized in the following.
We thus obtain the phase diagram shown in Fig. 5.
Notice that this phase diagram was corrrectly predicted in \cite{me99}.
Therein the correlation length for each phase was assumed to 
be given by the logarithm of the ratio of the largest 
and the second largest eigenvalue
of the matrix $C$. 

\renewcommand{\labelenumi}{5.\Roman{enumi}}
\begin{itemize}

\item Phase $C$
      ($\tilde{\alpha} > 1/2$ and $\tilde{\beta}>1/2$;
       $0<a,b<1$)

In this phase, the average density at bulk is $1/2$.
The average density decays near the right bounary as
\begin{equation}
  \label{density-max}
  \langle n_j \rangle_L
  =
  \frac12 
  -
  \frac{1}{2\sqrt{\pi} l^{\frac12}}.
\end{equation}
Here and in the following, we set $l=L-j+1$.
The density decays algebraically and hence the correlation length is 
infinite.
This average density profile is exactly the same as that for the 
totally asymmetric case.
The density decay at the left boundary can be obtained from 
the symmetry relation (\ref{ri-le}).

\item Phase $A_1$
      ($  \tilde{\alpha} < \tilde{\beta} 
        < \tilde{\alpha}/[(1-\tilde{\alpha})q+\tilde{\alpha}]$
       and $\tilde{\beta} < 1/2$;
       $aq<b<a$ and $b>1$ )

The average density at bulk is $\tilde{\alpha}$.
The density near the right boundary decays exponentially as
\begin{equation}
  \label{density-A1}
  \langle n_j \rangle_L
  =
  \tilde{\alpha}
  -
  \frac{(b^{-2}q,q;q)_{\infty}}
       {(a^{-1}b^{-1}q,ab^{-1}q;q)_{\infty}}
  \left[
   \frac{\tilde{\alpha}(1-\tilde{\alpha}) }
        {\tilde{\beta}(1-\tilde{\beta})}    
  \right]^l
  (1-2\tilde{\beta}),
\end{equation}
with the correlation length
\begin{equation}
\xi^{-1}
=
\ln \frac{\tilde{\beta}(1-\tilde{\beta})}
         {\tilde{\alpha}(1-\tilde{\alpha})}.
\end{equation}
On the other hand, the density near the left boundary takes
the constant value $\tilde{\alpha}$.
This fact is obtained by conbining the average density profile
near the right boundary for the high-density phase below and 
the symmetry (\ref{symmetry}).

\item Phase $A_2$
      ($q/(1+q) < \tilde{\alpha} < 1/2$ and $\tilde{\beta} > 1/2$;
       $1<a<q^{-1}$ and $b<1$)

In the bulk region, 
the average density takes the constant value $\tilde{\alpha}$.
On the other hand, the density profile near the right boundary decays
exponentially as
\begin{equation}
  \label{density-A2}
  \langle n_j \rangle_L
  =
  \tilde{\alpha}
  -
  \frac{(a-b)(1-ab)(abq,a^{-1}bq;q)_{\infty} (q;q)_{\infty}^4}
       {(a-1)^2(b-1)^2(aq,a^{-1}q,bq;q)_{\infty}}
  \frac{[4\tilde{\alpha}(1-\tilde{\alpha})]^l}
       {\sqrt{\pi} l^{\frac32}},
\end{equation}
with  
the correlation length 
\begin{equation}
\xi^{-1}
=
-\ln 4[\tilde{\alpha}(1-\tilde{\alpha})].
\end{equation}
But the decay is not purely exponential but with algebraic corrections.
At the left boundary, the density takes the constant value 
$\tilde{\alpha}$.

\item Phase $A_3$
     ( $\tilde{\beta} >  \tilde{\alpha}/[(1-\tilde{\alpha})q+\tilde{\alpha}]$
       and $\tilde{\alpha} < q/(1+q)$;
      $a>q^{-1}$ and $b<aq$)

The average density at bulk is $\tilde{\alpha}$.
Near the right boundary, the density decays exponentially as
\begin{equation}
  \label{density-A3}
  \langle n_j \rangle_L
  =
  \tilde{\alpha}
  -
  \frac{(1-ab)(1-(aq)^{-1})}
       {(1-b(aq)^{-1})(1+aq)}
  \left[\frac{(1+aq)(1+(aq)^{-1})}
             {(1+a)(1+a^{-1})}    \right]^l
\end{equation}
with the correlation length
\begin{equation}
\xi^{-1}
=
\ln \frac{q}
         {[\tilde{\alpha}+(1-\tilde{\alpha})q]^2}.
\end{equation}
This density profile has no correspondense for the totally
asymmetric case.
At the left boundary, the density takes the constant value 
$\tilde{\alpha}$.

\item Phase $B_1$ 
      ($ \tilde{\alpha}q/[(1-\tilde{\alpha})+q\tilde{\alpha}]  
        < \tilde{\beta} < \tilde{\beta} $
        and $\tilde{\alpha} < 1/2$;
      $bq<a<b$ and $a>1$)

The average density at bulk is $1-\tilde{\beta}$.
As can be seen from the calculation in Appendix,
The density takes the constant value near the right boundary.
\begin{equation}
  \label{density-B}
  \langle n_j \rangle_L
  =
  1-\tilde{\beta}
\end{equation}
Since this phase is related to the phase $A_1$ through the
symmetry (\ref{symmetry}),
the denisty decays exponentially near the left boundary 
with the correlation length
\begin{equation}
\xi^{-1}
=
\ln \frac{\tilde{\alpha}(1-\tilde{\alpha})}
         {\tilde{\beta}(1-\tilde{\beta})}.
\end{equation}

\item Phase $B_2$ 
     ($q/(1+q) < \tilde{\beta} < 1/2$ and $\tilde{\alpha} > 1/2$;
      $a<1$ and $1<b<q^{-1}$)

This phase is symmetric to phase $A_2$ through the symmetry 
(\ref{symmetry}).
The average density at bulk and near the right boundary is $1-\tilde{\beta}$.
Near the left boundary, the density decays exponentially 
with the correlation length
\begin{equation}
\xi^{-1}
=
-\ln 4[\tilde{\beta}(1-\tilde{\beta})].
\end{equation}

\item Phase $B_3$ ($a<bq$ and $b>q^{-1}$)
      ($\tilde{\beta} <  \tilde{\alpha}q/[1-\tilde{\alpha}+\tilde{\alpha}q]$
       and $\tilde{\beta} < q/(1+q)$;
       $a<bq$ and $b>q^{-1}$)

This phase is symmetric to phase $A_3$ through the symmetry 
(\ref{symmetry}).
The average density at bulk and near the right boundary is $1-\tilde{\beta}$.
The density decays exponentially with the correlation length
\begin{equation}
\xi^{-1}
=
\ln \frac{q}
         {[\tilde{\beta}+(1-\tilde{\beta})q]^2},
\end{equation}
near the left boundary.

\item Coexistence line
      ($\tilde{\alpha}=\tilde{\beta}<1/2$;
       $a=b>1$)

The average density shows linear profile at bulk,
\begin{equation}
  \label{density-coex}
  \langle n_j \rangle_L
  =
  \tilde{\alpha}
  +
  (1-2\tilde{\alpha})\frac{j}{L}.
\end{equation}
This is essentially the same as the result for the totally 
asymmetric case.

\end{itemize}

Before closing the section, we present a simulation 
result for the correlation length to show the 
differece between the phase $A_2$ and $A_3$.
The simulation was done on  $\tilde{\beta}=1$ line. 
For the totally asymmetric case, the system is in the 
$A_2$ phase on this line. The the correlation
length is given by $\xi_{A_2}=1/\ln4[\alpha(1-\alpha)]$.
On the other hand, for the partially asymmetric case,
the system is in the $A_3$ phase when $\tilde{\alpha}<q/(1+q)$.
The correlation length is given by 
$\xi_{A_3}=1/\ln\frac{q}{[\tilde{\beta}+(1-\tilde{\beta})q]^2}$.
The differences between these two expressions become
large especially when $q$ and $\alpha$ are small.
Especially, as $\alpha \rightarrow 0$, $\xi_{A_2}$ goes to zero
whereas $\xi_{A_3}$ goes to $-1/\ln q$.
In Fig. 6 the simulation result for the correlation 
length is shown for $p_R=1.0,\,P_L=0.9,\,\beta=10$ which corresponds
to $q=0.9,\,\tilde{\beta}=1$. It is clear that the 
the correlatin lenght approaches the finite value as 
$\alpha \rightarrow 0$ and is well described by the formula
for $\xi_{A_3}$.

\Section{Concluding Remarks}
\label{conc}
In this article, 
we have computed the average density profile of 
the partially asymmetric simple 
exclusion process with open boundaries.
The calculation has been done for a wide rage of parameters 
satisfying $0<p_L<p_R$ and $\alpha>0,\beta>0$.
The phase diagram for the correlation length has been obtained.
It has turned out that the phase diagram was correctly predicted in
the earlier paper \cite{me99}. 
In \cite{me99}, the phase diagram was obtained by assuming that
the correlation length is given by
the logarithm of the ratio of the largest and the second
largest eigenvalues of the matrix $C$.
The discussions were only for the phases with the exponentially
decaying profile.
In this article, we have not only confirmed this fact but also
obtained the asymptotic expressions of the average density profile
for all phases.

There are two key facts which allowed us to calculate 
the average density profile exactly in the thermodaynamic limit. 
One is that the commuation relation of the matrices $D,E$ becomes
a simple diagonal matrix and the other
is the formula (\ref{Mehler}) of the $q$-Hermite polynomials.

There seems to be many possible applications and 
generalizations of the analysis of this article.
First, it is possible to generalize the analysis in this paper to 
the partially asymmetric exclusion process 
on a ring with a single defect particle \cite{Mallick,Jafa}.
The corresponding totally asymmetric case was already solved in
\cite{Mallick}.
Second, the case where $p_L>p_R$ is also interesting.
Although the current was evaluated in  \cite{BECE},
more exact results are desirable.
Third, it would be interesting to apply the simialr analysis
to the multi-species models
\cite{EFGM,EKKM,AHR98-1,ADR,MMR}.
Compared to the ASEP,
much less is known about these models.
Several investigatios are now in progress \cite{RSS}.
The results about these will be reported elsewhere.

\section*{Acknowledgment}
The author would like to thank P. Deift, E. R. Speer 
and N. Rajewsky for fruitful discussions and comments. 
He also thanks the continuous encouragement of 
M. Wadati.
The author is a Research Fellow of the Japan Society
for the Promotion of Science.

\appendix
\renewcommand{\thesection}{Appendix \Alph{section}}
\renewcommand{\theequation}{\Alph{section}.\arabic{equation}}

\section{Evaluation of Integral $I_1$}
In this appendix, the integral $I_1$,
\begin{align}
I_1
&=
\frac14 
(ab;q)_{\infty}(q;q)_{\infty}^3
\int\frac{\d z_1}{2\pi i z_1}
\int\frac{\d z_2}{2\pi i z_2}
\notag\\
&\quad
\frac{(z_1^2,z_1^{-2},z_2^2,z_2^{-2};q)_{\infty}
      [(1+z_1)(1+z_1^{-1})]^{L-1}}
     {(az_1,a z_1^{-1},qz_1 z_2, q z_1^{-1} z_2^{-1},
       q z_1 z_2^{-1},q z_1^{-1} z_2^{-1}, bz_2,b z_2^{-1};q)_{\infty}
      (z_2 + z_2^{-1} -z_1-z_1^{-1})},
\label{I1def}
\end{align}
is evaluated.
First, for the case where $a,b<1$, 
both of the contours of $z_1$ and $z_2$ are unit circles.
We have
\begin{gather}
  I_1
  =
  I_1^{(0)}
  =
  \int_0^{\pi}
  \frac{\d \theta}{2\pi}
  \int_0^{\pi}
  \frac{\d \varphi}{2\pi}
  \frac{
        (e^{2i\theta},e^{-2i\theta},e^{2i\varphi},e^{-2i\varphi};q)_{\infty}
       }
       {(ae^{i\theta},        ae^{-i\theta},
         qe^{i(\theta+\varphi)}, qe^{-i(\theta+\varphi)}, 
         qe^{i(\theta-\varphi)}, qe^{-i(\theta-\varphi)}, 
         be^{i\varphi},        be^{-i\varphi};q)_{\infty} 
       }
  \notag\\
  \times
  \frac{ [2(1+\cos\theta)]^{L-1}  }
       { 2\cos\varphi - 2 \cos\theta }.  
  \label{I10}
\end{gather}
Second,  consider the case where $a$ becomes larger one 
but $b$ is still smaller than one. 
We assume 
\begin{equation}
  \label{na}
a>aq>aq^2>\cdots >aq^{n^{(a)}} >1> aq^{n^{(a)}+1}>\cdots . 
\end{equation}
Then the contour of $z_2$ is still a unit circle but the 
contour of $z_1$ has to be modified
to include all poles at $z_1=aq^k$ ($k=0,1,\ldots,n^{(a)}$)
and to exclude all
poles at $z_1=(a q^k)^{-1}$ ($k=0,1,\ldots,n^{(a)}$). 
Seperating the contributions from poles at $z_1=aq^{k}$
and $z_1=(aq^k)^{-1}$, the integral $I_1$ can be
rewritten as
\begin{align}
  \label{eq:I1-1}
I_1
&=
I_1^{(0)}
-
\frac{(ab;q)_{\infty} (q;q)_{\infty}^2}{2 a}
\sum_{k=0}^{n^{(a)}}
\frac{(-)^k q^{k(k-1)/2} 
      (a^2 q^{2k},a^{-2} q^{-2k};q)_{\infty}
      [\lambda_k^{(a)}]^{L-1} }
     {(q;q)_k (a^2 q^k;q)_{\infty}}
\notag\\
&\quad\times
\int_{C_0} \frac{\d z_2}{2\pi i z_2}
\frac{(z_2^2,z_2^{-2};q)_{\infty}}
     {(aq^{k+1}z_2,aq^{k+1}z_2^{-1},a^{-1}q^{-k}z_2,
       a^{-1}q^{-k}z_2^{-1},b z_2,bz_2^{-1};q)_{\infty}} . 
\end{align}
Here the contour $C_0$ denotes the unit circle.  
The $\lambda_k^{(c)}$'s are defined by
\begin{equation}
\lambda_k^{(c)}  
=
(1+c q^k)(1+c^{-1}q^{-k}),
\end{equation}
for $c=a,b$ and $k=0,1,2,\ldots$.
The intgral can be evaluated explicitly by using the general formula, 
\begin{equation}
  \label{int-aw}
  \int_C
  \frac{\d z}{2\pi i z}
  \frac{(z^2,z^{-2};q)_{\infty}}
       {(az,az^{-1},bz,bz^{-1},cz,cz^{-1},dz,dz^{-1};q)_{\infty}}
  =
  \frac{2(abcd;q)_{\infty}}
       {(q,ab,ac,ad,bc,bd,cd;q)_{\infty}}.
\end{equation} 
The contour $C$ is such that it includes all poles of the type
$f q^{k}$ and excludes all poles of the type $f^{-1}q^{-k}$
with $f=a,b,c,d$ and $k=0,1,2,\ldots$.
The parameters $a,b,c,d$ in this formula has nothing to do with 
the $a,b,c,d$ which appear in the rest of this article.
This formula plays a crutial role in proving the orthogonaliry
relation of the Askey-Wilson polynomials.  
The proof can be found in \cite{AW85}.
Now we get 
\begin{align}
  \label{I1a}
I_1
&=
I_1^{(0)}
+
I_1^{(a)},
\notag\\
I_1^{(a)}
&=
-\frac{(ab;q)_{\infty}}{a}
\sum_{k=0}^{n^{(a)}} 
\frac{(-)^k q^{k(k-1)/2} (a^2 q^{2k},a^{-2}q^{-2k};q)_{\infty}}
     {(q;q)_k (a^2 q^k,ab q^{k+1}, a^{-1}b q^{-k};q)_{\infty}}
[\lambda_k^{(a)}]^{L-1} . 
\end{align}
When $b$ also becomes larger than one,
there appear the terms which come from poles
at $z_2=bq^k$ and $z_2=(bq^k)^{-1}$ in (\ref{I1def}).
When (\ref{na}) and
\begin{equation}
  \label{nb}
b>bq>bq^2>\cdots >bq^{n^{(b)}} >1> bq^{n^{(b)}+1}>\cdots 
\end{equation} 
hold, we have
\begin{align}
\label{I1ab}
I_1
&=
I_1^{(0)}
+
I_1^{(a)}
+
I_1^{(b,0)}
+
I_1^{(b,b)}
+
I_1^{(b,a)},
\\
I_1^{(b,0)}
&=
\frac{(ab;q)_{\infty}(q;q)_{\infty}^2}{2 b}
\sum_{k=0}^{n^{(b)}}
\frac{(-)^k q^{k(k-1)/2} 
      (b^2q^{2k},b^{-2}q^{-2k};q)_{\infty} }
     {(q;q)_{k}(b^2q^k;q)_{\infty}} 
\notag\\
&\quad\times
\int_{C_0}\frac{\d z_1}{2\pi i z_1}
 \frac{(z_1^2,z_1^{-2};q)_{\infty} [(1+z_1)(1+z_1^{-1})]^{L-1}}
      {( az_1,az_1^{-1},
        bq^{k+1}z_1,bq^{k+1}z_1^{-1},
        b^{-1}q^{-k}z_1,b^{-1}q^{-k}z_1^{-1};q)_{\infty}} ,
\\
I_1^{(b,b)}
&=
-\sum_{k=0}^{n^{(b)}}
\sum_{m=k+1}^{n^{(b)}}
\frac{(-)^k q^{k(k-1)/2+(m-k)(m-k-1)/2}
      [\lambda_m^{(b)}]^{L-1} }
     {b (q;q)_k (q;q)_{m-k-1}}
\notag\\
&\quad\times
\frac{(q,ab,b^2q^{2k},b^{-2}q^{-2k},b^2q^{2m},b^{-2}q^{-2m};q)_{\infty}}
     {(q^{m-k},b^2 q^k,b^2q^{m+k+1},b^{-2}q^{-k-m},
       abq^m,ab^{-1}q^{-m};q)_{\infty} } ,
\\
I_1^{(b,a)}
&=
\sum_{k=0}^{n^{(b)}} 
\sum_{m:bq^k > aq^m}
\frac{(-)^{k+m} q^{k(k+1)/2+m(m+1)/2}
      [\lambda_m^{(a)}]^{L-1} }
     {b (q;q)_k (q;q)_{m}}
\notag\\
&\quad\times
\frac{(q,ab,b^2q^{2k},b^{-2}q^{-2k},a^2q^{2m},a^{-2}q^{-2m};q)_{\infty}}
     {(b^2 q^k,a^2 q^m,abq^{k+m+1},a^{-1}bq^{k+1-m},
       ab^{-1}q^{m-k},a^{-1}b^{-1}q^{-k-m};q)_{\infty}} .
\end{align}
Lastly, when $a<1$ and (\ref{nb}) holds, we have
\begin{equation}
  \label{I1b}
I_1
=
I_1^{(0)}
+
I_1^{(b,0)}
+
I_1^{(b,b)}.
\end{equation}

Now we turn to the calculation of the asymptotic expression of the 
integral $I_1$.

\begin{itemize}
\item{The case $a<1$ and $b<1$}

In this case, $I_1=I_1^{(0)}$.
We evaluate the asymptotic behavior of $I_1^{(0)}$ 
by employing the steepest decent method.
First we change the variable from $\theta,\varphi$ to
$u,y$ as 
\begin{gather}
\label{change1}
1+\cos\theta
=
2 e^{-u/L},
\\
\label{change2}
1+\cos\varphi
=
2 y e^{-u/L},
\end{gather}
to obtain
\begin{align}
\label{I1-1}
I_1
=
-\frac{4^{L+2}}{L^{\frac23}}
\int_{0}^{\infty} \d u u^{\frac12} e^{-u}
\sqrt{ \frac{1-e^{-u/L}}   
            {u/L       } } e^{-u/L}
\int_{0}^{e^{u/L}} \d y
\frac{ y^{\frac12}\sqrt{1-y e^{-u/L}} }
     {1-y}
\notag\\
\times
  \frac{
        (e^{2i\theta},e^{-2i\theta},e^{2i\varphi},e^{-2i\varphi};q)_{\infty}
       }
       {(ae^{i\theta},        ae^{-i\theta},
         qe^{i(\theta+\varphi)}, qe^{-i(\theta+\varphi)}, 
         qe^{i(\theta-\varphi)}, qe^{-i(\theta-\varphi)}, 
         be^{i\varphi},        be^{-i\varphi};q)_{\infty} 
       }.
\end{align}
Here $\theta$ and $\varphi$ are considerd as functions in
$u$ and $y$ through (\ref{change1}) and (\ref{change2}) respectively.
Now we can take the limit $L\rightarrow\infty$ in the integrand.
Changing the variable $y$ back to $\varphi$ by
\begin{equation}
1+\cos\varphi
=
2 y,
\end{equation}
we have
\begin{equation}
  I_1
  \simeq
  -\frac{\sqrt{\pi} (q;q)_{\infty}^2 4^{L+1} }
        {2 (a;q)_{\infty}^2 L^{\frac32} }
  \int_{0}^{\pi}
  \d \varphi
  \frac{(e^{2i\varphi},e^{-2i\varphi};q)_{\infty}}
       {(qe^{i\varphi},qe^{i\varphi},qe^{-i\varphi},qe^{-i\varphi},
         be^{i\varphi},be^{i\varphi};q)_{\infty}} .
\end{equation}
Using the formula (\ref{int-aw}),
we get
\begin{equation}
  I_1
 \simeq
  -\frac{\pi^{\frac32} (1-b) 4^{L+1} }
        {(a,b;q)_{\infty}^2 L^{\frac32}}.
\end{equation}

\item{The case $a<1$ and $b>1$}

The integral $I_1$ is given by (\ref{I1b}).
The main contributions come from $I_1^{(0)}$ and
$I_1^{(b,0)}$. Each contribution behaves as
$4^L$. whilest the normalization $Z_L$ behaves as
$[(1+b)(1+b^{-1})]^L$ for this case. 
Since $(1+b)(1+b^{-1})$ is larger than $4$, $I_1$ is negligible
compared to $Z_L$. 
So we do not compute the explicit expression.

\item{The case $a>1$}

The main contribution comes 
from the $k=0$ term in the summation of $I_1^{(a)}$.
We have
\begin{equation}
I_1
\simeq
-\frac{a(1-ab)(a^{-2};q)_{\infty}[(1+a)(1+a^{-1})]^{L-1}}
      {(a^{-1}b;q)_{\infty}}.
\end{equation}

\end{itemize}

\setcounter{equation}{0}
\renewcommand{\thesection}{Appendix \Alph{section}}
\section{Evaluation of Integral $I_2$}
In this Appendix, the integral $I_2$,
\begin{align}
I_2
&=
\frac14 
(ab;q)_{\infty}(q;q)_{\infty}^3
\int\frac{\d z_1}{2\pi i z_1}
\int\frac{\d z_2}{2\pi i z_2}
\notag\\
&\quad
\frac{(z_1^2,z_1^{-2},z_2^2,z_2^{-2};q)_{\infty}
      [(1+z_1)(1+z_1^{-1})]^{j-1}
      [(1+z_2)(1+z_2^{-1})]^{L-j} }
     {(az_1,a z_1^{-1},qz_1 z_2, a z_1^{-1} z_2^{-1},
       q z_1 z_2^{-1},q z_1^{-1} z_2^{-1}, bz_2,b z_2^{-1};q)_{\infty}
      (z_2 + z_2^{-1} -z_1-z_1^{-1})},
\label{I2def}
\end{align}
is evaluated.
First, for the case where $a,b<1$, 
both of the contours of $z_1$ and $z_2$ are unit circles.
We have 
\begin{gather}
  I_2
  =
  I_2^{(0)}
  =
  \int_0^{\pi}
  \frac{\d \theta}{2\pi}
  \int_0^{\pi}
  \frac{\d \varphi}{2\pi}
  \frac{
        (e^{2i\theta},e^{-2i\theta},e^{2i\varphi},e^{-2i\varphi};q)_{\infty}
       }
       {(ae^{i\theta},        ae^{-i\theta},
         qe^{i(\theta+\varphi)}, qe^{-i(\theta+\varphi)}, 
         qe^{i(\theta-\varphi)}, qe^{-i(\theta-\varphi)}, 
         be^{i\varphi},        be^{-i\varphi};q)_{\infty} 
       }
  \notag\\
  \times
  \frac{ [2(1+\cos\theta)]^{j-1}  [2(1+\cos\varphi)]^{L-j}  }
       { 2\cos\varphi - 2 \cos\theta } . 
  \label{I20}
\end{gather}
Since we will calculate the average density profile 
near the right boundary,
we set $l=L-j+1$. 
Similarly to the case of the integal $I_1$,
when $a$ or $b$ or both of them become larger than one,
there apper other contributions besides $I_2^{(0)}$.
When (\ref{na}) and (\ref{nb}) hold,
the result is
\begin{align}
I_2
&=
I_2^{(0)}
+
I_2^{(a,0)}
+
I_2^{(a,a)}
+
I_2^{(a,b)}
+
I_2^{(b,0)}
+
I_2^{(b,b)}
+
I_2^{(b,a)},
\label{I2all}
\\
I_2^{(a,0)}
&=
-\frac{(ab;q)_{\infty}(q;q)_{\infty}^2}{2 a}
\sum_{k=0}^{n^(a)}
\frac{(-)^k q^{k(k-1)/2} 
      (a^2q^{2k},a^{-2}q^{-2k};q)_{\infty}   
      [\lambda_k^{(a)}]^{L-l} }
     {(q;q)_{k}(a^2q^k;q)_{\infty}} , 
\notag\\
&\quad\times
\int_{C_0}\frac{\d z_2}{2\pi i z_2}
 \frac{(z_2^2,z_2^{-2};q)_{\infty} [(1+z_2)(1+z_2^{-1})]^{l-1}}
      {(aq^{k+1}z_2,aq^{k+1}z_2^{-1},
        a^{-1}q^{-k}z_2,a^{-1}q^{-k}z_2^{-1},
        bz_2,bz_2^{-1};q)_{\infty}} ,
\\
I_2^{(a,a)}
&=
\sum_{k=0}^{n^{(a)}}
\sum_{m=k+1}^{n^{(a)}}
\frac{(-)^k q^{k(k-1)/2+(m-k)(m-k-1)/2}
      [\lambda_k^{(a)}]^{L-l} [\lambda_m^{(a)}]^{l-l} }
     {a (q;q)_k (q;q)_{m-k-1}}
\notag\\
&\quad\times
\frac{(q,ab,a^2q^{2k},a^{-2}q^{-2k},a^2q^{2m},a^{-2}q^{-2m};q)_{\infty}}
     {(q^{m-k},a^2 q^k,a^2q^{m+k+1},a^{-2}q^{-k-m},
       abq^m,ba^{-1}q^{-m};q)_{\infty} } ,
\\
I_2^{(a,b)}
&=
-\sum_{k=0}^{n^{(a)}} 
\sum_{m:aq^k > bq^m}
\frac{(-)^{k+m} q^{k(k+1)/2+m(m+1)/2}
      [\lambda_k^{(a)}]^{L-l} [\lambda_m^{(b)}]^{l-l} }
     {a (q;q)_k (q;q)_{m}}
\notag\\
&\quad\times
\frac{(q,ab,a^2q^{2k},a^{-2}q^{-2k},b^2q^{2m},b^{-2}q^{-2m};q)_{\infty}}
     {(a^2 q^k,b^2 q^m,abq^{k+m+1},ab^{-1}q^{k+1-m},
       a^{-1}bq^{m-k},a^{-1}b^{-1}q^{-k-m};q)_{\infty}} ,
\\
I_2^{(b,0)}
&=
\frac{(ab;q)_{\infty}(q;q)_{\infty}^2}{2 b}
\sum_{k=0}^{n^{(b)}}
\frac{(-)^k q^{k(k-1)/2} 
      (b^2q^{2k},b^{-2}q^{-2k};q)_{\infty}   
      [\lambda_k^{(b)}]^{l-1} }
     {(q;q)_{k}(b^2q^k;q)_{\infty}} 
\notag\\
&\quad\times
\int_{C_0}\frac{\d z_1}{2\pi i z_1}
 \frac{(z_1^2,z_1^{-2};q)_{\infty} [(1+z_1)(1+z_1^{-1})]^{L-l}}
      {( az_1,az_1^{-1},
        bq^{k+1}z_1,bq^{k+1}z_1^{-1},
        b^{-1}q^{-k}z_1,b^{-1}q^{-k}z_1^{-1};q)_{\infty}} ,
\\
I_2^{(b,b)}
&=
-\sum_{k=0}^{n^{(b)}}
\sum_{m=k+1}^{n^{(b)}}
\frac{(-)^k q^{k(k-1)/2+(m-k)(m-k-1)/2}
      [\lambda_k^{(b)}]^{l-1} [\lambda_m^{(b)}]^{L-l} }
     {b (q;q)_k (q;q)_{m-k-1}}
\notag\\
&\quad\times
\frac{(q,ab,b^2q^{2k},b^{-2}q^{-2k},b^2q^{2m},b^{-2}q^{-2m};q)_{\infty}}
     {(q^{m-k},b^2 q^k,b^2q^{m+k+1},b^{-2}q^{-k-m},
       abq^m,ab^{-1}q^{-m};q)_{\infty} } ,
\\
I_2^{(b,a)}
&=
\sum_{k=0}^{n^{(b)}} 
\sum_{m:bq^k > aq^m}
\frac{(-)^{k+m} q^{k(k+1)/2+m(m+1)/2}
      [\lambda_k^{(b)}]^{l-1} [\lambda_m^{(a)}]^{L-l} }
     {b (q;q)_k (q;q)_{m}}
\notag\\
&\quad\times
\frac{(q,ab,b^2q^{2k},b^{-2}q^{-2k},a^2q^{2m},a^{-2}q^{-2m};q)_{\infty}}
     {(b^2 q^k,a^2 q^m,abq^{k+m+1},a^{-1}bq^{k+1-m},
       ab^{-1}q^{m-k},a^{-1}b^{-1}q^{-k-m};q)_{\infty}} .
\end{align}

Now we consider the asymptotic expression of the integral $I_2$.
We take the limit $L\rightarrow\infty$ at first and then 
take the limit $l\rightarrow\infty$.

\begin{itemize}
\item{The case $a,b<1$}

In this case, $I_2=I_2^{(0)}$. 
The evaluation for this case proceeds analogously to 
the evaluation of $I_1^{(0)}$.
Changing the variables $\theta,\varphi$ to $u,y$ as 
in (\ref{change1}) and (\ref{change2}),
we get 
\begin{align}
\label{I2-1}
I_2
=
-\frac{4^{L+2}}{L^{\frac23}}
\int_{0}^{\infty} \d u u^{\frac12} e^{-u}
\sqrt{ \frac{1-e^{-u/L}}   
            {u/L       } } e^{-u/L}
\int_{0}^{e^{u/L}} \d y
\frac{ y^{l-\frac12}\sqrt{1-y e^{-u/L}} }
     {1-y}
\notag\\
\times
  \frac{
        (e^{2i\theta},e^{-2i\theta},e^{2i\varphi},e^{-2i\varphi};q)_{\infty}
       }
       {(ae^{i\theta},        ae^{-i\theta},
         qe^{i(\theta+\varphi)}, qe^{-i(\theta+\varphi)}, 
         qe^{i(\theta-\varphi)}, qe^{-i(\theta-\varphi)}, 
         be^{i\varphi},        be^{-i\varphi};q)_{\infty} 
       }.
\end{align}
We take the limit $L\rightarrow\infty$ in this expression and 
change the varialbe $y$ back to $\varphi$
to get
\begin{equation}
I_2
  \simeq
  -\frac{\sqrt{\pi} (q;q)_{\infty}^2 4^{L+2} }
        {2 (a;q)_{\infty}^2 L^{\frac23}}
  \int_{0}^{\pi}
  \d \phi
  \left[ \frac12 (1+\cos\varphi) \right]^{l}
  \frac{(qe^{2i\varphi},qe^{-2i\varphi};q)_{\infty}}
       {(qe^{i\varphi},qe^{i\varphi},qe^{-i\varphi},qe^{-i\varphi},
         be^{i\varphi},be^{i\varphi};q)_{\infty}}. 
\end{equation}
We can consider the limit $l\rightarrow\infty$ by 
using the steepest descent method.
We have
\begin{equation}
  I_2
  \simeq
  -\frac{ 2 \cdot 4^{L+1}\pi }
        {(a,b;q)_{\infty}^2 L^{\frac23} l^{\frac12}}.
\end{equation}

\item{The case $1<a<q^{-1}$ and $b<1$}

In this case, the integral $I_2$ is given by 
$I_2
=
I_2^{(0)}+I_2^{(a,0)}$.
The main contribution comes from the $k=0$ term in $I_2^{(a,0)}$.
\begin{align}
  \label{I2a0}
I_2
&\simeq
-(ab,a^{-2};q)_{\infty} (q;q)_{\infty}^2 [(1+a)(1+a^{-1})]^{L-l}   
\notag\\
&\quad\times
\int_{C_0} \frac{\d z_2}{2\pi i z_2}
\frac{(z_2^2,z_2^{-2};q)_{\infty} [(1+z_2)(1+z_2^{-1})]^{l-1}}
     {(a^{-1}z_2,a^{-1}z_2^{-1},aqz_2,
       aqz_2^{-1},bz_2,bz_2^{-1};q)_{\infty}}.
\end{align}
Taking the limit $l\rightarrow\infty$ by using the steeptest
descent method, we get
\begin{equation}
I_2
\simeq
-\frac{4^l (ab,a^{-2};q)_{\infty}(q;q)_{\infty}^2
      [(1+a)(1+a^{-1})]^{L-l}  }
      {a \sqrt{\pi} l^{\frac32} (a^{-1},qa,b;)_{\infty}^2}.
\end{equation}

\item{The case $a>q^{-1}$ and $aq>b$}

In general, the integral for this case has 
the expression (\ref{I2all}).
The main contribution comes from the 
$k=0,m=1$ term in $I_2^{(a,a)}$.
We have
\begin{equation}
  \label{I2aa}
I_2
\simeq
-\frac{(1-ab)(1-a^{-2}q^{-2})(a^{-2};q)_{\infty}
       [(1+a)(1+a^{-1})]^{L-l} 
       [(1+aq)(1+a^{-1}q^{-1})]^{l-1}}
      {(a^{-1}bq^{-1};q)_{\infty}}.     
\end{equation}

\item{The case $a>b>aq$ and $b>1$}

The main contribution for this case comes from the 
$k=0,m=1$ term in $I_2^{(a,b)}$.
We have
\begin{equation}
  \label{I2ab}
I_2
\simeq
-\frac{(1-ab)(q,a^{-2},b^{-2};q)_{\infty} 
       [(1+a)(1+a^{-1})]^{L-l}
       [(1+b)(1+b^{-1})]^{l-1} }
      {a (a^{-1}b,a^{-1}b^{-1},ab^{-1}q;q)_{\infty}} .
\end{equation}

\item{The case $b>1$ and $b>a$}

In this case, the normalization $Z_L$ behaves as $[(1+b)(1+b^{-1})]^L$.
All contributions to $I_2$ can be neglected compared to $Z_L$.
Hence we do not compute the asymptotic expression explicitly for this case.
\end{itemize}

\newpage
\begin{large}
\noindent
Figure Captions
\end{large}

\vspace{10mm}
\noindent
Fig. 1 : One-dimensional partially 
asymmetric simple exclusion process with open
boundaries. Particles have hard-core exclusion interaction and 
tend to hop to the right (resp. left) nearest neiboring
site with rate $p_R$ (resp. $p_L$). There are also particle
injection (resp. ejection) at the left (resp. right) edge.

\vspace{10mm}
\noindent
Fig. 2 : Space-time diagram of the ASEP from Monte-Carlo simulations. 
The holizontal axis represents the site number $j$ 
whereas the vertical axis represents time.
The existence of pariticle is represented as a black point.
The lattice length is taken to be $L=200$.
The bulk hopping rates are taken to be $p_R=1,\,p_L=0$.
After some transient time, the system practically goes to a 
steady state. The steady state depends crutially on the values
of the boundary parameters.

\vspace{10mm}
\noindent
Fig. 3 : 
Average density profile of the ASEP from Monte-Carlo simulations. 
The holizontal axis represents the site number $j$ 
whereas the vertical axis represents the average density.
The lattice length is taken to be $L=200$.
The bulk hopping rates are taken to be $p_R=1,\,p_L=0$

\vspace{10mm}
\noindent
Fig. 4 : The phase diagram of the current.
Regions $A,B$ and $C$ are called the low-density phase, 
the high-density phase and the maximal current phase
respectively.

\vspace{10mm}
\noindent
Fig. 5 : The phase diagram of the correlation length 
in the $\tilde{\alpha}$-$\tilde{\beta}$ plane 
for the partially asymmetric case.
The low-density phase 
(resp. high-density phase) is divided into
three phases, $A_1,A_2$ and $A_3$ (resp. $B_1,B_2$ and $B_3$).

\vspace{10mm}
\noindent
Fig. 6 : The correlation length $\xi$ for the 
case $p_R=1,p_L=0.9,\tilde{\beta}=1$. The solid line is 
the theoretical prediction given by
$\xi=1/\ln\frac{q}{[\tilde{\beta}+(1-\tilde{\beta})q]^2}$
whereas the black dots are the simulation data.

\newpage 
\vspace*{5cm}
\begin{figure}[h!]
\scalebox{0.5}{\includegraphics{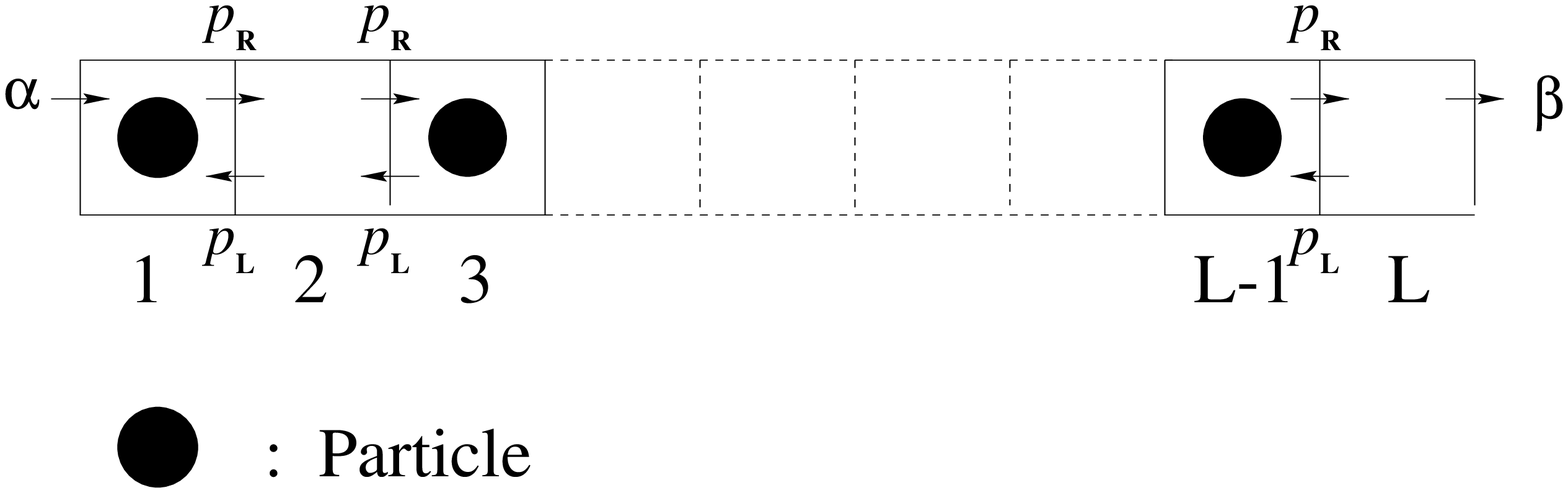}}
\end{figure}

\newpage
\vspace*{1cm}
\begin{picture}(500,500)

\put(20,305){\scalebox{0.9}[0.35]{\includegraphics{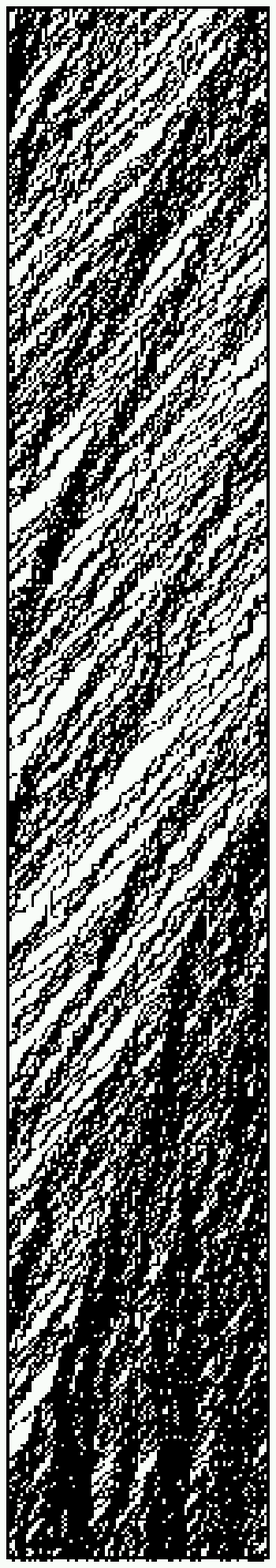}}}
\put(50,295){\vector(1,0){50}}
\put(70,280){$j$}
\put(15,380){\vector(0,1){70}}
\put(-3,410){\rotatebox{90}{$t$}}
\put(20,265){($A$) low density phase}
\put(32,253){$p_R=1.0, \,p_L=0.0$}
\put(37,240){$\alpha=0.2,\,\beta=1.0$}

\put(220,308){\scalebox{0.9}[0.35]{\includegraphics{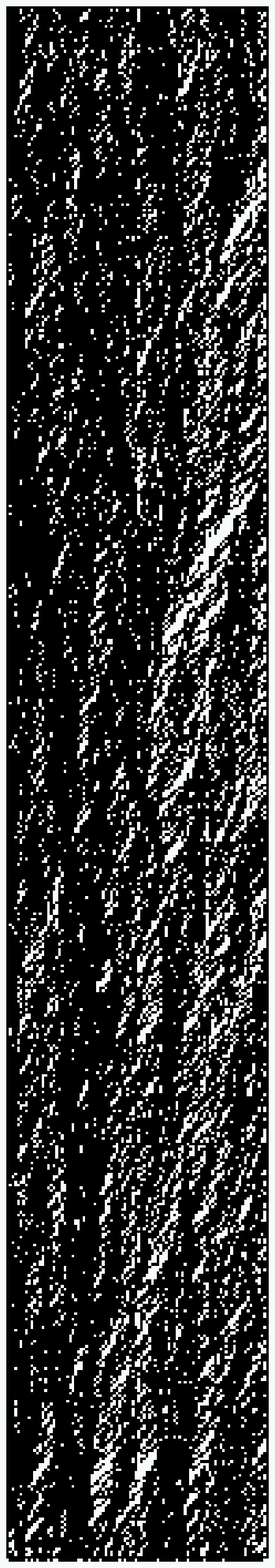}}}
\put(250,295){\vector(1,0){50}}
\put(270,280){$j$}
\put(215,380){\vector(0,1){70}}
\put(197,410){\rotatebox{90}{$t$}}
\put(200,265){($C$) maximal current phase}
\put(232,253){$p_R=1.0, \,p_L=0.0$}
\put(237,240){$\alpha=1.0,\,\beta=1.0$}

\put(20,5){\scalebox{0.9}[0.35]{\includegraphics{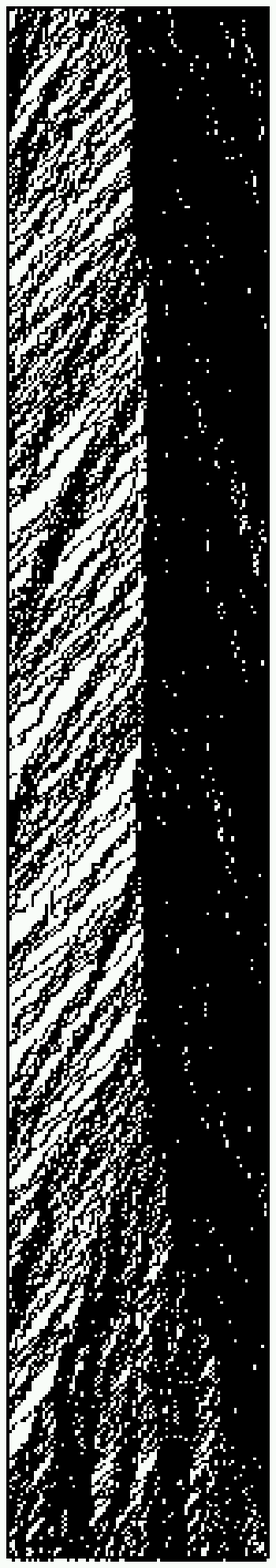}}}
\put(50,-5){\vector(1,0){50}}
\put(70,-20){$j$}
\put(15,80){\vector(0,1){70}}
\put(-3,110){\rotatebox{90}{$t$}}
\put(40,-35){coexistence line}
\put(32,-47){$p_R=1.0, \,p_L=0.0$}
\put(37,-60){$\alpha=0.2,\,\beta=0.2$}

\put(220,0){\scalebox{0.9}[0.35]{\includegraphics{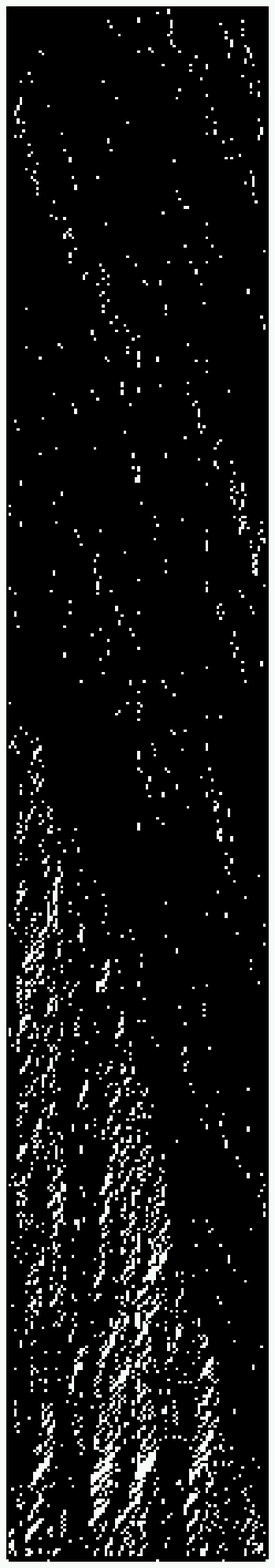}}}
\put(250,-5){\vector(1,0){50}}
\put(270,-20){$j$}
\put(215,80){\vector(0,1){70}}
\put(197,110){\rotatebox{90}{$t$}}
\put(220,-35){($B$) high density phase}
\put(232,-47){$p_R=1.0, \,p_L=0.0$}
\put(237,-60){$\alpha=1.0,\,\beta=0.2$}

\end{picture}

\newpage
\begin{picture}(500,300)

\put(0,190){\scalebox{0.6}[0.6]{\includegraphics{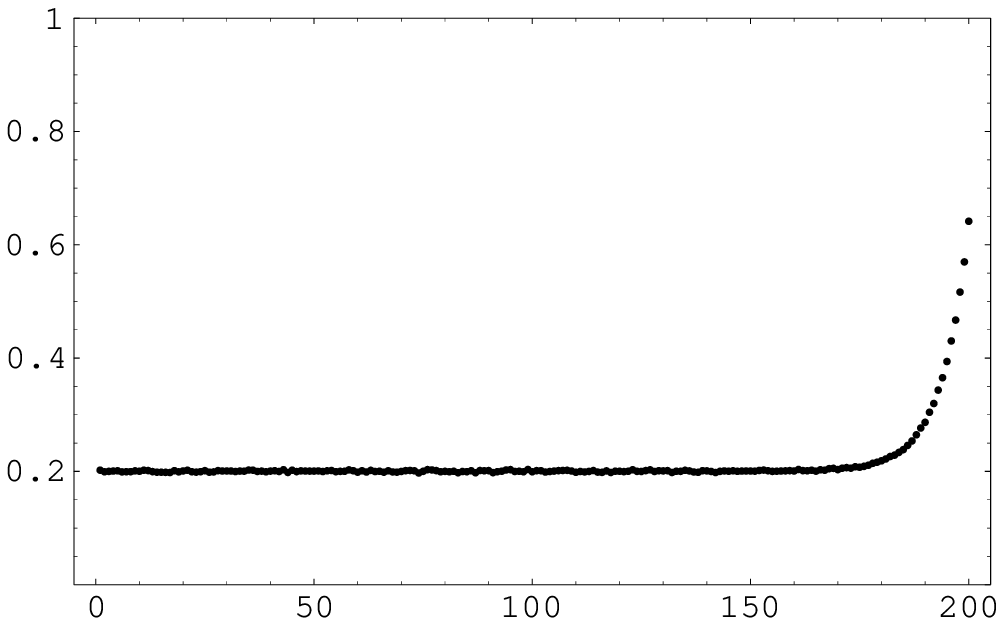}}}
\put(70,182){\vector(1,0){50}}
\put(90,170){$j$}
\put(-3,212){\vector(0,1){70}}
\put(-20,240){\rotatebox{90}{$\langle n_j \rangle$}}
\put(30,155){($A$) low density phase}
\put(42,140){$p_R=1.0, \,p_L=0.0$}
\put(47,125){$\alpha=0.2,\,\beta=0.25$}

\put(200,190){\scalebox{0.6}[0.6]{\includegraphics{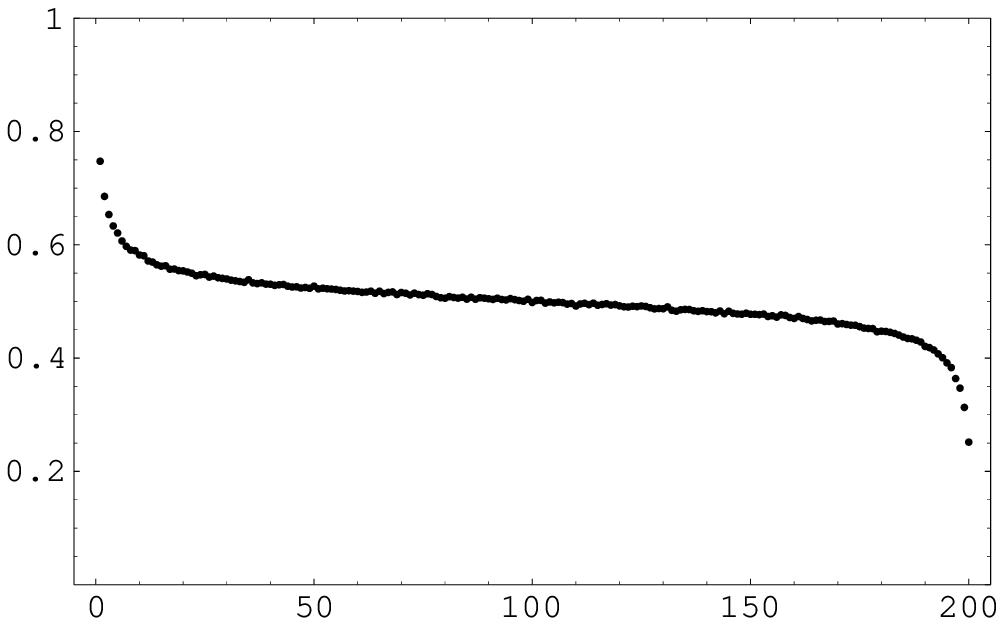}}}
\put(270,182){\vector(1,0){50}}
\put(290,170){$j$}
\put(220,155){($C$) maximal current phase}
\put(242,140){$p_R=1.0, \,p_L=0.0$}
\put(247,125){$\alpha=1.0,\,\beta=1.0$}

\put(0,0){\scalebox{0.6}[0.6]{\includegraphics{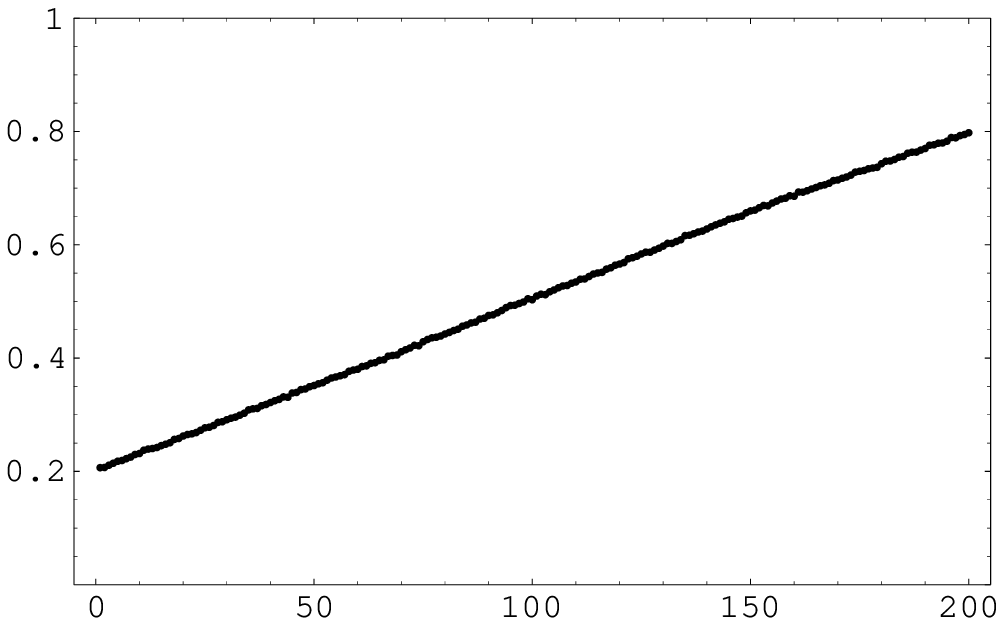}}}
\put(70,-5){\vector(1,0){50}}
\put(90,-20){$j$}
\put(50,-35){coexistence line}
\put(42,-50){$p_R=1.0, \,p_L=0.0$}
\put(47,-65){$\alpha=0.2,\,\beta=0.2$}

\put(200,0){\scalebox{0.6}[0.6]{\includegraphics{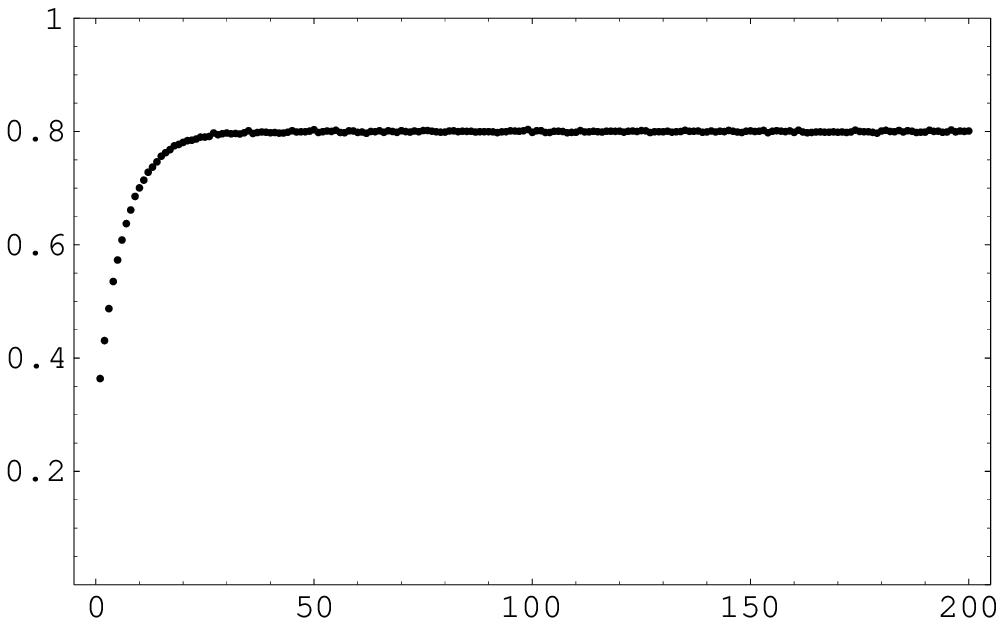}}}
\put(270,-5){\vector(1,0){50}}
\put(290,-20){$j$}
\put(230,-35){($B$) high density phase}
\put(242,-50){$p_R=1.0, \,p_L=0.0$}
\put(247,-65){$\alpha=0.25,\,\beta=0.2$}

\end{picture}

\newpage
\vspace*{3cm}
\begin{figure}
\scalebox{1.0}{\includegraphics{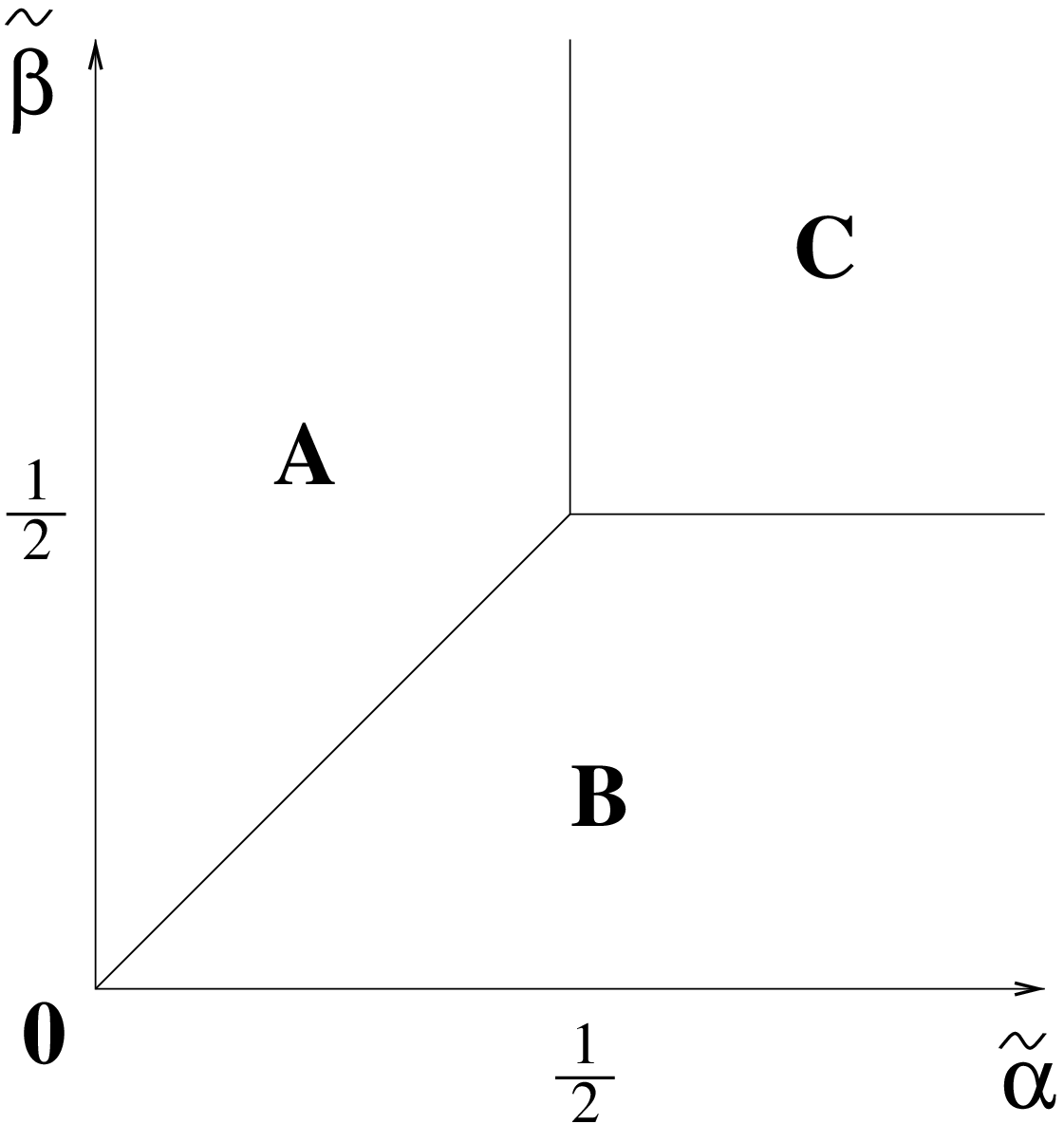}}
\end{figure}

\newpage
\vspace*{3cm}
\begin{figure}
\scalebox{1.0}{\includegraphics{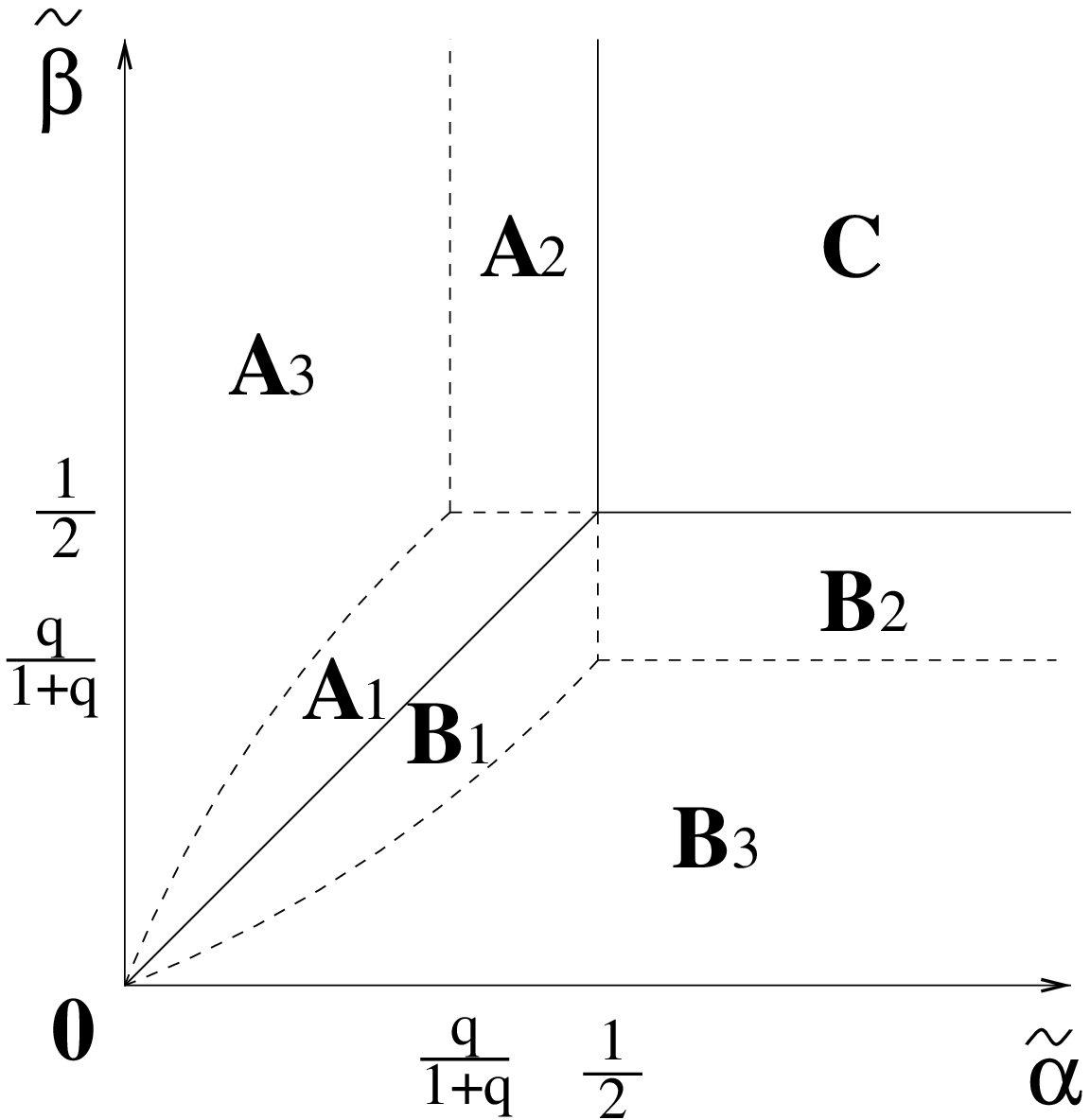}}
\end{figure}

\newpage
\vspace*{3cm}
\begin{picture}(500,300)

\put(0,100){\scalebox{1.0}[1.0]{\includegraphics{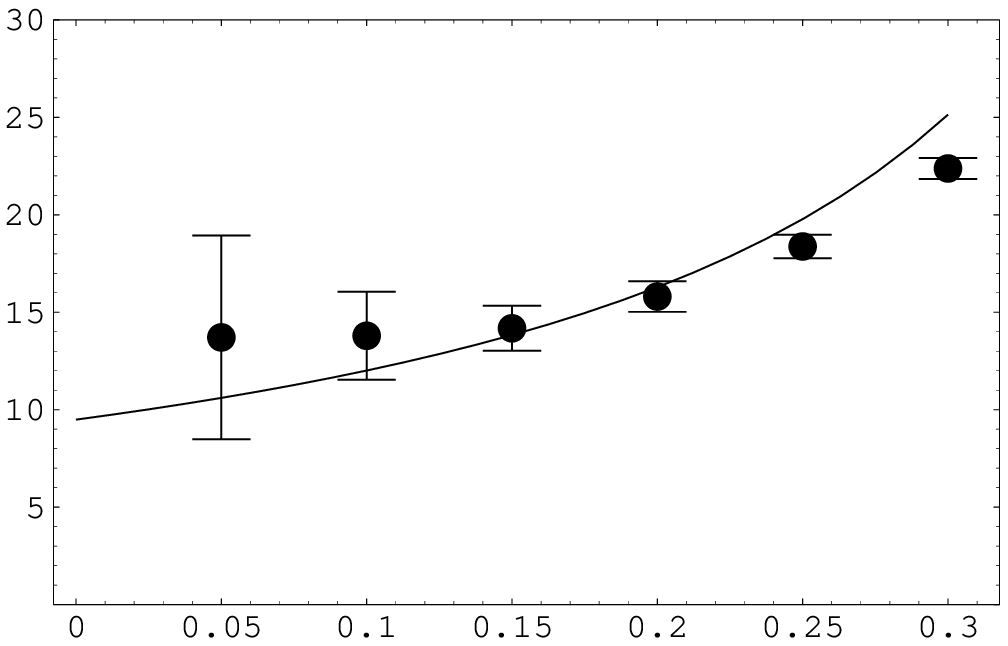}}}
\put(-20,270){\scalebox{1.5}{$\xi$}}
\put(280,85){\scalebox{1.5}{$\tilde{\alpha}$}}

\end{picture}

\end{document}